\documentclass[twocolumn,prl,amsmath,amssymb]{revtex4}
\usepackage{graphicx}
\usepackage{epstopdf}
\usepackage{amsmath}
\usepackage{color}
\begin{document}

\title{Ultra-strong nonlinear optical processes and trigonal warping in MoS$_2$ layers}
\author{A. S\"ayn\"atjoki$^{1,2,*}$, L. Karvonen$^{1,*}$, H. Rostami$^{3,*}$, A. Autere$^1$, S. Mehravar$^4$, A. Lombardo$^5$, R. A. Norwood$^4$, T. Hasan$^5$, N. Peyghambarian$^{1,2,4}$, H. Lipsanen$^1$, K. Kieu$^4$, A. C. Ferrari$^5$, M. Polini$^{3}$, Z. Sun$^1$}
\affiliation{$^1$Aalto University, Department of Micro and Nanosciences,Tietotie 3, FI-02150 Espoo, Finland}
\affiliation{$^2$University of Eastern Finland, Institute of Photonics, Yliopistokatu 7, FI-80100 Joensuu, Finland}
\affiliation{$^3$Istituto Italiano di Tecnologia, Graphene Labs, Via Morego 30, I-16163 Genova, Italy}
\affiliation{$^4$University of Arizona, College of Optical Sciences, 1630 EUniversity Blvd, Tucson, AZ 85721, USA}
\affiliation{$^5$Cambridge Graphene Centre, University of Cambridge, Cambridge CB3 0FA, UK}
\begin{abstract}
We report ultra-strong high-order nonlinear multiphoton processes in monolayer MoS$_2$ (1L-MoS$_2$): the third harmonic is thirty times stronger than the second harmonic, and the fourth harmonic is comparable to the second harmonic. We find that second and third harmonic processes are strongly dependent on elliptical polarization, which can be used to selectively tune harmonic generation with different orders. We explain this by calculating the nonlinear response functions of 1L-MoS$_2$ with a continuum-model Hamiltonian and quantum-mechanical diagrammatic perturbation theory, highlighting the crucial role of trigonal warping. A similar effect is expected for all other transition-metal dichalcogenides. Our results pave the way for efficient and tunable harmonic generation based on layered materials for various applications, including microscopy and imaging.
\end{abstract}
\maketitle
\footnote{*These authors contributed equally to this work}
Nonlinear optical processes, such as harmonic generation\cite{boyd}, are of great interest for various applications, e.g. microscopy\cite{Zipfel_nb_03,Bhawalkar_rpp_96}, therapy\cite{Zipfel_nb_03,Bhawalkar_rpp_96}, frequency conversion\cite{boyd,Broderick_JOSAB_02} and data storage\cite{Bhawalkar_rpp_96}. Nonlinear optical phenomena can generate high-energy photons by converting $n=2,3,4,\dots$ low-energy photons into one high-energy photon. These are usually referred to as second-, third- and fourth-harmonic generation (SHG, THG and FHG)\cite{boyd,Zipfel_nb_03,Bhawalkar_rpp_96,Broderick_JOSAB_02}. Due to different selection rules\cite{boyd,Pavone_13}, various harmonic processes are distinct from optically-pumped laser phenomena (e.g. optically-pumped amplification\cite{Saleh}), and other typical single-photon processes (e.g. single-photon excited photoluminescence\cite{boyd}), in which the energy of the generated photons is smaller than the pump photons. Therefore, multiphoton harmonic processes have been widely exploited for various applications (e.g. all-optical signal processing in telecommunications\cite{boyd,Willner_JLT14}, medicine\cite{Zipfel_nb_03,Bhawalkar_rpp_96}, and data storage\cite{Bhawalkar_rpp_96}), as well as to study various transitions forbidden under low-energy single-photon excitation\cite{Zipfel_nb_03,Bhawalkar_rpp_96}. The physical origin of these processes is the nonlinear polarization induced by an electromagnetic field ${\bm E}$. This gives rise to higher harmonic components, the $n$-th harmonic component amplitude being proportional to $|{\bm E}|^n$ \cite{boyd}. Quantum mechanically, higher-harmonic generation consists in the annihilation of $n$ pump photons and generation of a photon with $n$ times the pump energy. Because an $n$-th order nonlinear optical process requires $n$ photons to be present simultaneously, the probability for higher-order processes is lower than for lower order\cite{boyd}. Thus, higher-order processes are typically weaker and require higher intensities\cite{zhu_sci_1997,tsang_pra_1995}.

Graphene and related materials (GRMs) are at the center of an ever increasing research effort due to their unique and complementary properties, making them appealing for a wide range of photonic and optoelectronic applications\cite{Bonaccorso_np_10,Butler_an_13,KoppensNN,Ferrari_ns,Wang_nn_12,Xu_np_14,Sun_np_16}. Amongst these, semiconducting transition-metal dichalcogenides (TMDs) are of particular interest due to their direct bandgap when in monolayer form\cite{Mak_prl_10,Splendiani_nl_2010}, leading to an increase in luminescence efficient by a few orders of magnitude compared with the bulk material\cite{Mak_prl_10,Splendiani_nl_2010,Eda_nl_2011,Goki_an_13,Amani_sci_15,berraquero_arxiv_2016}. 1L-MoS$_2$ has a single layer of Mo atoms sandwiched between two layers of S atoms in a trigonal prismatic lattice. Therefore, in contrast to graphene, it is non-centrosymmetric and belongs to the space group $D_{3h}^1$\cite{Li_nl_13}. The lack of spatial inversion symmetry makes 1L-MoS$_2$ an interesting material for nonlinear optics, since second-order nonlinear processes are present only in non-centrosymmetric materials\cite{boyd}. However, when stacked, MoS$_2$ layers are arranged mirrored with respect to one another\cite{Li_nl_13}, therefore MoS$_2$ with an even number of layers (EN) is centrosymmetric and belongs to the $D_{3d}^3$ space group\cite{Li_nl_13}, producing no second harmonic (SH) signal. On the other hand, MoS$_2$ with any odd number of layers (ON) is non-centrosymmetric. SHG from 1L-MoS$_2$ has already been experimentally demonstrated by several groups\cite{Kumar_prb_13,Wang_an_13_1,Malard_prb_13,Li_nl_13,Wang_ami_14,Trolle_prb_14,Clark_prb_14,Bonaccorso_ome_14,Seyler_nn_15}.

Here, we present experimental and theoretical work on nonlinear harmonic generation in 1L and few-layer (FL) MoS$_2$ flakes. We report ultra-strong THG and FHG from 1L-MoS$_2$. In comparison to SHG, the THG is more than one order of magnitude larger and FHG has the same magnitude as SHG. This is surprising, since one normally expects the intensity of non-linear optical processes to decreases with $n$\cite{boyd,Pavone_13}. One therefore expects the SHG intensity to be much larger than THG and FHG, although even-order processes only exist in non-centrosymmetric materials. Our results show that this expectation is wrong in the case of 1L-MoS$_2$. The point is that, at sufficiently low photon-frequencies (in our experiments the photon energy of the pump is $0.8{\rm eV}$), SHG only probes the low-energy band structure of 1L-MoS$_2$. This is nearly rotationally invariant\cite{Kuc_prb_11, Kadantsev_ssc_12, Wang_nn_12, Shi_prb_13, Zahid_aip_13,Kormanyos_prb_13, Qiu_prl_13, gibertini_prb_2014}, but with corrections due to trigonal warping. It is because of these corrections\cite{Kadantsev_ssc_12, Zahid_aip_13,Kormanyos_prb_13}, fully compatible with the $D_{3h}^1$ space group\cite{boyd}, but reducing the full rotational symmetry of the low-energy bands to a three-fold rotational symmetry\cite{boyd}, that a finite amplitude of non-linear harmonic processes with even $n$ can exist at low photon energies. Thus, lack of spatial inversion symmetry is only a necessary but not sufficient condition for the occurrence of SHG. We demonstrate that the observed THG/SHG intensity ratio can be explained by quantum mechanical calculations based on finite-temperature many-body diagrammatic perturbation theory\cite{rostami_prb_2016} and low-energy continuum-model Hamiltonians that include trigonal warping\cite{RG15}. We show that these nonlinear processes are sensitive to the number of layers, their symmetry, relative orientation, and the elliptical polarization of the excitation light. Similar effects are expected for all other TMDs. This paves the way for the assembly of heterostructures with tailored nonlinear properties.
\begin{figure}
\centerline{\includegraphics[width=90mm]{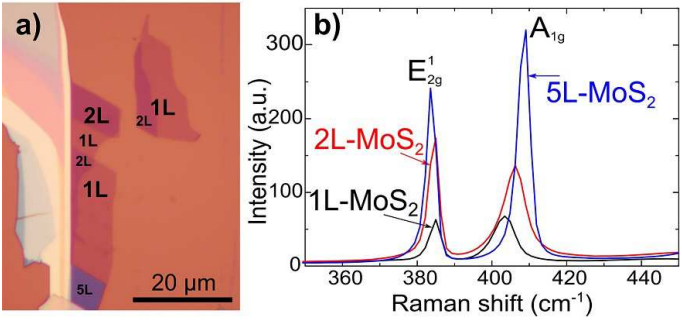}}
\caption{\label{Raman_PL} a) Optical micrograph with single-layer, bilayer, and five-layer areas marked by 1L, 2L and 5L, respectively. b) Raman spectra of the same sample.}
\end{figure}

MoS$_2$ flakes are produced by micromechanical cleavage (MC) of bulk MoS$_2$\cite{Novoselov_PNAS05,Bonaccorso_mt_12} onto Si+285nm SiO$_2$ substrate. 1L-MoS$_2$ and bilayer (2L-MoS$_2$) samples are identified by a combination of optical contrast\cite{SundNL13,Casiraghi_NL07} and Raman spectroscopy\cite{LeeACSN4,ZhangPRB87}. Raman spectra are acquired by a Renishaw micro-Raman spectrometer equipped with a 600 line/mm grating and coupled with an Ar$^+$ ion laser at 514.5nm. Fig.\ref{Raman_PL} shows the MoS$_2$ flakes studied in this work and their Raman signatures. A reference MC graphene sample is also prepared on a similar substrate.

\begin{figure}
\centerline{\includegraphics[width=80mm]{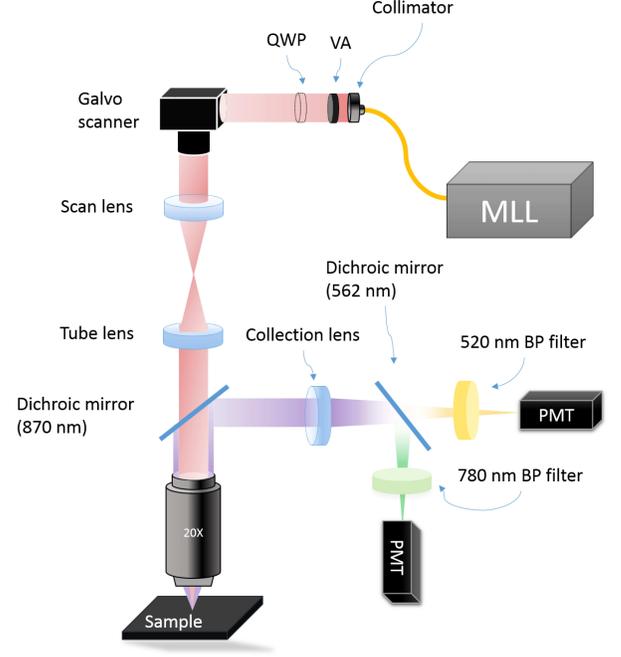}}
\caption{\label{fig:MPMschema} Schematic diagram of the multiphoton microscope. MLL: linearly polarized mode-locked fiber laser. VA: variable attenuator. QWP: quarter-wave plate. QWP is inserted only when we study the dependence of SHG and THG on the elliptical polarization of the pump light. BP filter: Bandpass filter. PMT: Photomultiplier tube.}
\end{figure}

Nonlinear optical measurements are carried out with the setup of Fig.\ref{fig:MPMschema}\cite{saynatjoki13,kieu10}. As excitation source, we use an erbium doped mode-locked fiber laser with a $\sim$50MHz repetition rate, maximum average power$\sim$60mW and pulse duration$\sim$150fs, which yields an estimated pulse peak power$\sim8$kW\cite{kieu10high}. The laser beam is scanned with a galvo mirror and focused on the sample using a microscope objective. The back-scattered second and third harmonic signals are split into different branches using a dichroic mirror and then detected using photomultiplier tubes (PMTs). For two-channel detection, the light is split into two PMTs using a dichroic mirror with 560nm cut-off. After the dichroic mirror, the detected wavelength range can be further refined using bandpass filters. The light can also be directed to a spectrometer (OceanOptics QE Pro-FL) to analyze the spectral properties of the generated light. The average power on sample is kept between 10 and 28mW with a typical measurement time$\sim$5$\mu$s, which prevents sample damage and enables high signal-to-noise-ratio, even with acquisition time per pixel in the $\mu s$ range.

\begin{figure}
\centerline{\includegraphics[width=80mm]{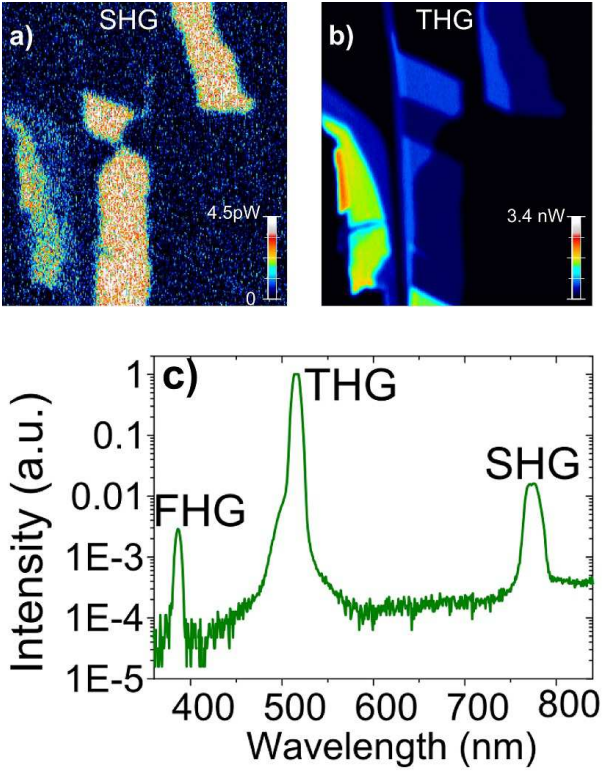}}
\caption{\label{fig:1040_MPM_a} a) SHG and b) THG map of the MoS$_2$ flake in Fig.\ref{Raman_PL}a). c) Optical spectrum of the nonlinear signal from 1L-MoS$_2$ with a peak irradiance$\sim30~{\rm GW/cm}^2$.}
\end{figure}

SHG and THG images of the MoS$_2$ sample are shown in Figs.\ref{fig:1040_MPM_a}a,b). The SHG signal is generated in 1L-MoS$_2$, while 2L-MoS$_2$ appears dark. As discussed above, the second-order nonlinear response is present in 1L-MoS$_2$, which is non-centrosymmetric. However, when stacked to form 2L-MoS$_2$, MoS$_2$ layers are mirrored one with respect to another\cite{Li_nl_13,Kumar_prb_13}. Therefore, EN-MoS$_2$ is centrosymmetric \cite{Li_nl_13,Kumar_prb_13}, and belongs to the $D_{3d}^3$ space group\cite{Li_nl_13,Kumar_prb_13}, producing no SHG signal. On the other hand, ON-MoS$_2$ is non-centrosymmetric \cite{Li_nl_13,Kumar_prb_13}.

We note that strong THG is detected compared with SHG, even for 1L-MoS$_2$, as shown in Fig.\ref{fig:1040_MPM_a}b). THG was previously reported for a thick ($N\geq 10$) MoS$_2$ flake\cite{Wang_ami_14}, but here we see it down to 1L-MoS$_{2}$. However, THG is not observed from the thickest areas of our flake, with N$\>$30, as in Ref.\cite{Wang_ami_14}. The output spectrum in Fig.\ref{fig:1040_MPM_a}c) further confirms that we are observing SHG and THG together. Peaks for THG and SHG at$\sim$520 and$\sim$780nm can be seen, as well as a peak at$\sim$390nm, corresponding to a four-photon process. This is detected only in 1L-MoS$_2$. Its intensity is$\sim$5.5 times lower than SHG, and two orders of magnitude smaller than THG.
\begin{figure}
\centerline{\includegraphics[width=80mm]{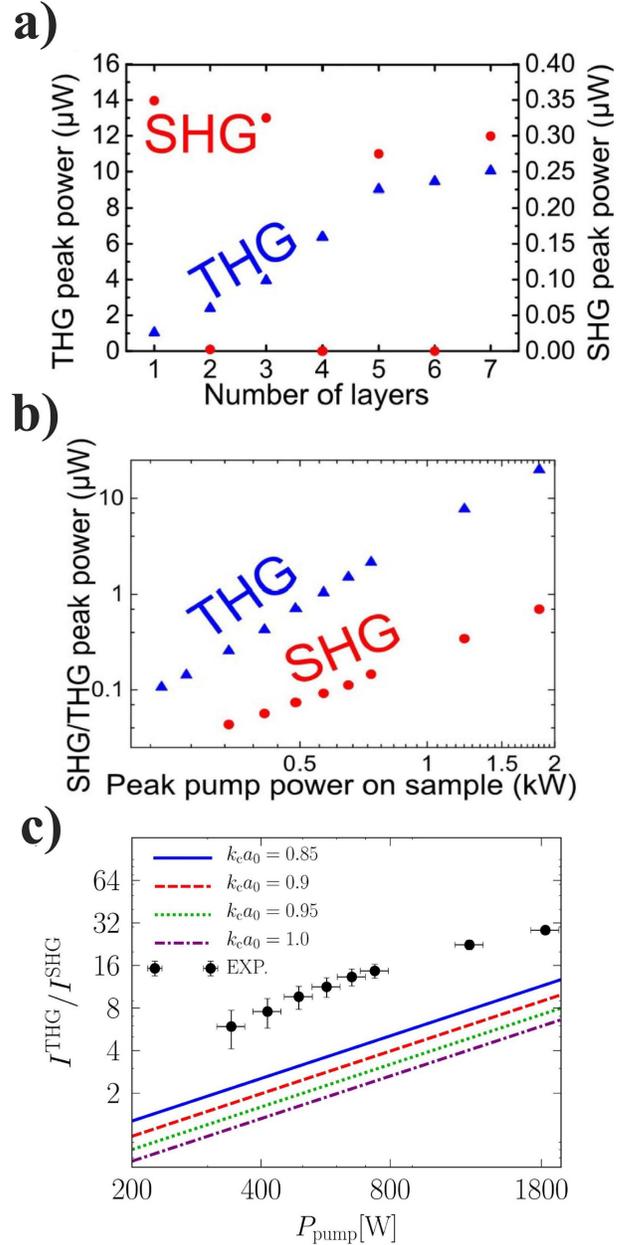}}
\caption{\label{fig:1040_MPM_b} a) SHG and THG intensities as functions of N. b) Power dependence of SHG and THG in 1L-MoS$_2$. c) Experimental and theoretical THG/SHG irradiance ratio as a function of $P_{\rm pump}$. Different theoretical curves refer to different values of the ultra-violet cut-off $k_{\rm c}$ (measured in units of $1/a_{0}=\sqrt{3}/a$ where $a \approx 3.16~{\text \AA}$ is the lattice constant of 1L-MoS$_2$ \cite{Mak_prl_10,Splendiani_nl_2010}).}
\end{figure}

SHG signals on areas with N=3, 5, 7 have nearly the same intensity as 1L-MoS$_{2}$, Fig.\ref{fig:1040_MPM_b}a). This contrasts Ref.\cite{Li_nl_13}, where a pump laser at 810nm was used. We attribute this difference to the fact that photons generated in the second-order nonlinear process in our setup with a 1560nm pump wavelength have an energy$\sim$1.6eV (780nm), below the band gap of 1L-MoS$_2$ \cite{Mak_prl_10,Splendiani_nl_2010}, therefore not adsorbed, unlike the SHG signal in Ref.\cite{Li_nl_13}.

Based on the measured SHG and THG intensities, we can estimate the nonlinear susceptibilities $\chi^{(2)}$ and $\chi^{(3)}$. $\chi^{(2)}$ can be calculated from the measured average powers of the fundamental and SH signals as follows\cite{Janisch_Scirep_14}:
\begin{equation}\label{eq: x2}
\chi^{(2)}_s=\sqrt{\frac{\epsilon_0 c \lambda_2^4 P_{2\omega} R \tau^2 (n_2+1)^2(n_1+1)^2}{32{\rm NA}^2\tau_2 P_{\rm pump} \phi}}~,
\end{equation}
where $\tau$ is the pulse width, $P_{\rm pump}$ is the average power of the incident fundamental (pump) beam and $P_{2\omega}$ stands for the generated SH beam power, $R$ is the repetition rate, NA=0.5 is the numerical aperture, $\lambda_2$=780nm is the SH wavelength, $\tau=\tau_{2}=150$ fs are the pulse durations at fundamental and SH wavelengths, $\phi=8\pi\int_{0}^{1}|\cos^{-1}\rho-\rho\sqrt{1-\rho^2}|^2 \rho ~ \mathrm{d}\rho=3.56$ from Ref.\cite{Janisch_Scirep_14}, and $n_1=n_2= \sim 1.45$ are the refractive indices of the substrate at the wavelengths of the fundamental and SHG, respectively. The effective bulk-like second order susceptibility of MoS$_2$ ($\chi^{(2)}_{\text{eff}}$) can be obtained from Eq.\ref{eq: x2} with $\chi^{(2)}_{\text{eff}}=\frac{\chi_s^{(2)}}{t_{\text{MoS}_2}}$, where $t_{\text{MoS}_2}=0.75$nm is the 1L-MoS$_{2}$ thickness\cite{Ferrari_ns,Wang_nn_12}. We obtain the effective second order susceptibility $\chi^{(2)}_{\text{eff}}\sim~2.2~{\rm pm}/{\rm V}$ for 1L-MoS$_2$.
The third-order susceptibility $\chi^{(3)}$ of MoS$_2$ is estimated by comparing the measured THG signal from MoS$_2$ to that of 1L-graphene (SLG):
\begin{equation}\label{equation2}
\chi^{(3)} \approx\frac{t_{\rm gr}}{t_{\text{MoS}_2}}\sqrt{\frac{\text{THG}_{\text{MoS}_2}}{\text{THG}_{\rm gr}}}\chi^{(3)}_{\rm gr}~.
\end{equation}
With $t_{\rm gr}\sim$0.33nm the SLG thickness, and THG$_{\rm gr}$ and THG$_{{\rm MoS}_2}$ the measured signals from SLG and MoS$_2$, respectively. Using $\chi^{(3)}_{\rm gr}\sim 3\times 10^{-7}~{\rm esu}\sim4.2\times 10^{-15}~{\rm m^2/V^2}$\cite{saynatjoki13}, we find $\chi^{(3)}\sim 2.8\times 10^{-7}~{\rm esu}\sim3.9\times 10^{-15}~{\rm m^2/V^2}$, comparable to that of SLG in the same frequency range that we used in our experiment. This is remarkable, as SLG is known to have a large $\chi^{(3)}$\cite{hendry_prl_2010,kumar_prb_2013}, 2 orders of magnitude larger than that of bulk glass\cite{hong_prx_2013} and$\sim$5 times larger than gold\cite{hong_prx_2013}. Furthermore, MoS$_2$ is transparent at this telecommunication wavelength due to its$\sim$1.9eV gap\cite{Mak_prl_10,Splendiani_nl_2010,Goki_an_13}, while SLG absorbs 2.3\% of the light\cite{Ferrari_ns,Nair_s_08}. Therefore, MoS$_2$ and possibly other TMDs are promising for integration with optical waveguides or fibers for all-optical nonlinear devices, where materials with nonlinear properties are essential, such as all-optical modulators and signal processing devices\cite{Sun_np_16}.

The SHG and THG  power dependence follows quadratic and cubic trends, respectively, Fig.\ref{fig:1040_MPM_b}b). At the power levels of our measurements, THG is up to 30 times stronger than SHG. We attribute such a large THG/SHG ratio to the approximate rotational invariance of the MoS$_2$ band structure at low energies, which is broken by trigonal warping. Fig.\ref{fig:1040_MPM_b}c) plots the THG/SHG ratio obtained from the experiments and microscopic calculations based on the ${\bm k} \cdot {\bm p}$ theory\cite{RG15} and finite-temperature diagrammatic perturbation theory\cite{rostami_prb_2016} (details in Methods). The calculations are factor of two smaller than the experimental data. Considering the complexity of the investigated non-linear optical processes and the fact that our calculations ignore high-energy band structure effects\cite{gibertini_prb_2014} and many-body renormalizations\cite{Gruning_prb_14}, we believe this to be a satisfactory agreement, indicating the importance of trigonal warping in harmonic generation.
\begin{figure}
\centerline{\includegraphics[width=65mm]{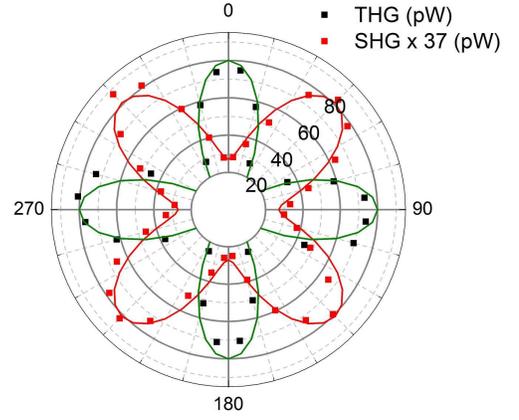}}
\caption{\label{fig:1560_poldep} Dependence of SHG and THG intensities on the elliptical polarization of the pump light in 1L-MoS$_2$. The polar plot angle corresponds to linearly polarized light when $\theta=0^{\circ}+m\cdot90^{\circ}$, and gives circularly polarized pump light when $\theta=45^{\circ}+m\cdot90^{\circ}$. The SHG power is multiplied by a factor of 37 to fit in the same scale as THG.}
\end{figure}
\begin{figure*}
\centerline{\includegraphics[width=150mm]{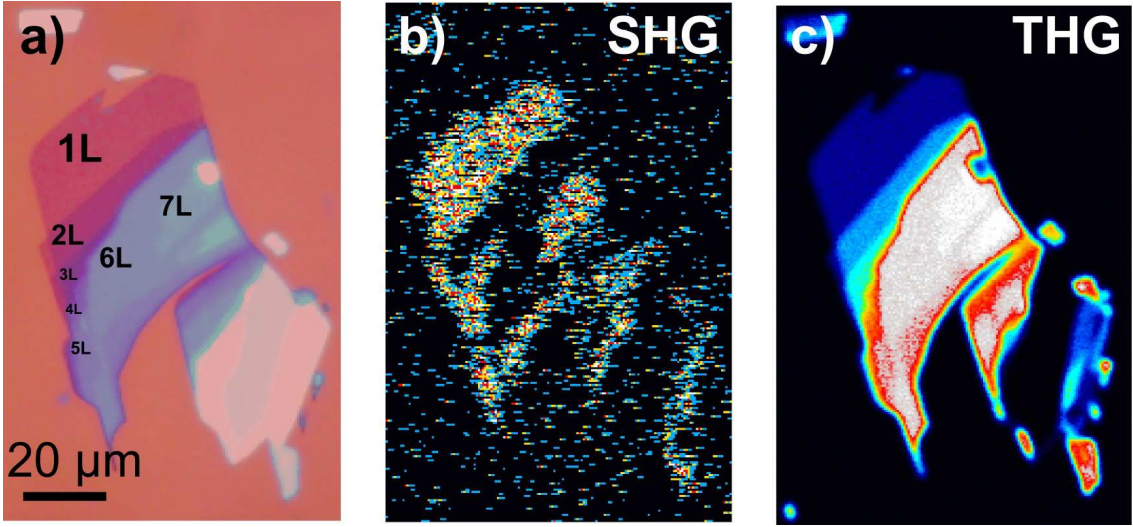}}
\caption{\label{fig:1560_sec}a) Optical micrograph, b) SHG and c) THG images of flake with few-layer areas under 1560nm excitation.}
\end{figure*}

FHG generally derives from cascades of lower-order nonlinear multi-photon processes\cite{pfister_ol_1997,zhu_sci_1997}. With an excitation wavelength of 1560nm, this could be, e.g., a cascade of two SHG processes, where 780nm photons are first generated through SHG ($\omega_{1560 \rm nm}+\omega_{1560\rm nm}\Rightarrow \omega_{780\rm nm}$) and then undergo another SHG process ($\omega_{780\rm nm}+\omega_{780\rm nm}\Rightarrow \omega_{390\rm nm}$). To yield a FHG at $390~{\rm nm}$ of the same intensity as SHG at 780nm in this cascaded process, one would need a conversion efficiency (defined as $P_{2\omega}/P_{\rm pump}$ \cite{boyd}) for the second SHG process (i.e., $\omega_{780\rm nm}+\omega_{780\rm nm}\Rightarrow \omega_{390\rm nm}$) to be close to unity. However, we observe a conversion efficiency$\sim10^{-10}$ for SHG. Therefore, we conclude that our FHG does not arise from cascaded SHGs. Another possible cascade process is based on THG ($\omega_{1560\rm nm}+\omega_{1560\rm nm}+\omega_{1560\rm nm}\Rightarrow \omega_{520\rm nm}$) and sum-frequency generation ($\omega_{520\rm nm}+\omega_{1560\rm nm}\Rightarrow \omega_{390\rm nm}$). We find that THG strongly increases up to N=5, as for Fig.\ref{fig:1040_MPM_b}a). Therefore, we expect this cascaded process to have a similar trend with N. However, we only observe FHG in 1L-MoS$_2$. Thus, we can also exclude this cascade process, and conclude that we observe a direct $\chi^{(4)}$ process in 1L-MoS$_2$. To the best of our knowledge, this is the first observation of FHG in a monolayer GRM.

We now discuss the dependence of our results on the elliptical polarization of the incident light. We consider an incident laser beam with arbitrary polarization, i.e.~${\bm E}=|{\bm E}|{\hat {\bm \varepsilon}}_{\pm}$ with ${\hat {\bm \varepsilon}}_{\pm} = \hat {\bm x} \cos(\theta)  \pm i \hat{\bm y} \sin(\theta)$. Using the crystal symmetries of 1L-MoS$_2$, we derive (see Methods) the following expressions for the second- and third-order polarizations ${\bm P}^{(2)}$ and ${\bm P}^{(3)}$:
\begin{equation}\label{eq:secondorder-circ}
{\bm P}^{(2)}= \epsilon_0 \chi^{(2)}_{yyy} |\bm E|^2 [ \mp i  \sin(2\theta) \hat {\bm x} -  \hat {\bm y}  ]
\end{equation}
and
\begin{equation}\label{eq:thirdorder-circ}
{\bm P}^{(3)}=\epsilon_0   \chi^{(3)}_{yyyy} |\bm E|^3 \hat{\bm \varepsilon}_{\pm}  \cos(2\theta)~.
\end{equation}
Note that $\theta=0^{\circ}$ corresponds to a linearly polarized laser along the $\hat{\bm x}$ direction which is perpendicular to the mirror symmetry plane of $D^1_{3h}$ symmetry group, while $\theta=45^{\circ}$ corresponds to a circularly polarized laser. From Eq.\ref{eq:secondorder-circ} we expect the intensity of SHG in response to a circularly polarized pump laser to be twice that of a linearly polarized laser. Eq.\ref{eq:thirdorder-circ} implies vanishing THG in response to a circularly polarized pump laser.

We thus measure the SHG and THG dependence on elliptical polarization using a linearly polarized laser and a rotating quarter-wave plate (QWP). Depending on the angle $\theta$ between the QWP axes and the polarization, the excitation light will be linearly ($\theta=0^{\circ}+m\cdot90^{\circ}$) or circularly ($\theta=45^{\circ}+m\cdot90^{\circ}$) polarized. Fig.\ref{fig:1560_poldep} shows that the experiments are in excellent agreement with Eqs.\ref{eq:secondorder-circ},\ref{eq:thirdorder-circ}. The THG signal is maximum for a linearly polarized excitation laser, while it vanishes for circularly polarized excitation. Note that SHG is always visible, but its intensity is maximum for circularly polarized light.

Given that harmonic generation is strongly dependent on the symmetry and stacking of layers and different monolayer TMDs (e.g. WSe$_2$,MoSe$_2$), all have similar nonlinear response\cite{Li_nl_13,Kumar_prb_13,Seyler_nn_15,Sun_np_16}, one could use heterostructures (e.g. MoS$_2$/WSe$_2$) to engineer SHG and other nonlinear processes for high photon-conversion efficiency for a wide range of applications requiring the generation of higher frequencies. This may lead to the use of layered materials and heterostructures for applications utilizing optical nonlinearities (e.g. all-optical devices, frequency combs, high-order harmonic generation, multiphoton microscopy and therapy etc.).
\section{Methods}
\subsection{Determination of MoS$_2$ thickness from SHG and THG signals}
SHG and THG for few-layer MoS$_2$ ($N$=1...7) are studied on the flake in Fig.\ref{fig:1560_sec}a. SHG and THG images are shown in Figs.\ref{fig:1560_sec}b,c. At $1560~{\rm nm}$, the contrast between 1 and 3L areas is small, as well as the contrast between 3-, 5- and 7L regions (Fig.\ref{fig:1560_sec}b).

The THG signal increases up to N=7, Figs.\ref{fig:1040_MPM_b}a,\ref{fig:1560_sec}c. On the other hand, the SHG signal (Fig.\ref{fig:1560_sec}b) is only generated in ON areas, due to symmetry\cite{Li_nl_13}. Therefore, the areas that have intensity between the 3-, 5- and 7L areas in Fig.\ref{fig:1560_sec}c but appear dark in SHG, can be identified as 4 and 6L. The dependence of the intensities of THG and SHG on N is plotted in Fig.\ref{fig:1040_MPM_b}a). Thus, the combination of SHG and THG can be used to accurately identify N at least up to 7. The THG signal develops as a function of N. Using Maxwell's equations for a non-linear medium with thickness $t$ and considering the slowly varying amplitude approximation\cite{boyd,Butcher_and_Cotter}, we obtain:
\begin{equation}
\frac{I_{3\omega}}{I_{\rm in}} \approx \frac { (3\omega)^2 I^2_{\rm in} }{16 n^3_1 n_3 \epsilon^2_0 c^4} \left |\chi^{(3)} (-3\omega;\omega,\omega,\omega)\right |^2 t^2 {\rm sinc}^2 \left ( \frac{\Delta k t}{2} \right ),
\end{equation}
where $I_{\rm in}$ and $I_{3\omega}$ are the intensity of the incident and THG light, respectively and $\chi^{(3)} (-3\omega;\omega,\omega,\omega)$ is the third order optical susceptibility. Note that $n_{j=1,3} =\sqrt{\epsilon^{(1)}(j \omega )}$ in which $\epsilon^{(1)}$ is the linear dielectric function of multilayer TMD. $\Delta k t$ is the phase mismatch between the fundamental and third harmonic generated waves. For $\Delta k t \approx 0$, THG adds up quadratically with light propagation length (i.e. $t \propto N$). The signal starts to saturate for N=6. The possible reasons for sub-quadratic signal build-up can be either phase mismatch, or absorption\cite{Eda_nl_2011}. For THG, $\Delta k= 3k_{\rm in} \pm k_{3\omega}$, where $k_{\rm in}$ and $k_{3\omega}$ are the wavevectors of the incident and THG signals, respectively, where the plus sign indicates THG generated in the backward direction, while minus identifies forward generated THG. Even for backward generated THG, $\Delta k t \approx 0$ for 6L-MoS$_2$ ($\sim 4.3~{\rm nm}$\cite{Radisavljevic_nn_11}). This rules out phase mismatch as the origin of the signal saturation when $N \leq 6$. Therefore we assume that the signal saturation is due to absorption of the third harmonic light.
\subsection{Continuum-model Hamiltonian and current matrix elements for 1L-MoS$_2$}
For 1L-MoS$_2$ we use the low-energy ${\bm k} \cdot {\bm p}$ continuum-model Hamiltonian described in Ref.\cite{RG15}. Around the ${\rm K}$ and ${\rm K}^\prime$ points the model Hamiltonian contains isotropic ${\cal H}_{\rm i}$ and trigonal warping ${\cal H}_{\rm tw}$ contributions, i.e. ${\cal H}={\cal H}_{\rm i}+{\cal H}_{\rm tw}$, with:
\begin{eqnarray}\label{eq:H_iso}
{\cal H}_{\rm i}({\bm  k},\tau, s) &=& \frac{\lambda_0\tau s}{2}+\frac{\Delta+\lambda\tau s}{2}\sigma_z +t_0 a_0 {\bm  k}\cdot {\bm \sigma}_\tau
\nonumber \\
&+&\frac{\hbar^2|{\bm  k}|^2}{4m_0}(\alpha+\beta\sigma_z)~,
\end{eqnarray}
and
\begin{eqnarray}\label{eq:H_tw}
{\cal H}_{\rm tw}({\bm  k},\tau, s) &=& t_1 a_0^2({\bm  k} \cdot {\bm \sigma}^{\ast}_{\tau})
\sigma_x ({\bm  k}\cdot{\bm \sigma}^{\ast}_{\tau})
\nonumber\\
&+& t_2 a_0^3 \tau (k_x^3-3k_xk_y^2)(\alpha'+\beta'\sigma_z)~.
\end{eqnarray}
Here, $s=\pm$ is a spin index, $\tau=\pm$ is a valley index, and ${\bm \sigma}_{\tau}=(\tau\sigma_x,\sigma_y)$, with $\sigma_x$ and $\sigma_y$ ordinary $2\times 2$ Pauli matrices operating on a suitable conduction/valence band basis\cite{RG15}. We note that the terms in the Hamiltonian that contain the parameters $\Delta$, $\beta$, $\beta'$ and $\lambda_0$ are related to broken spatial inversion symmetry in 1L-MoS$_2$. The trigonal warping term contains three parameters, $\alpha'$,$\beta'$, and $t_1$. The contribution to the band dispersion due to trigonal warping has the characteristic form $z_{\pm}\cos{(3\phi)}$, where $z_{\pm}=t_2(\alpha'\pm\beta')\pm4 t_0 t_1/\left[2\Delta-(\lambda_0-\lambda)\tau s\right]$, and $z_+$ ($z_-$) stands for conduction (valence) band\cite{RG15_2}. According to {\it ab}-{\it initio} calculations\cite{Zahid_aip_13,Kormanyos_prb_13}, symmetry considerations\cite{Kormanyos_prb_13,RMA13}, and experimental evidence\cite{AH14}, the valence band of 1L-MoS$_2$ is strongly warped, while the conduction band is nearly isotropic.

The Hamiltonian ${\cal H}$ can be diagonalized. Eigenvalues $\epsilon^{{\rm c} ({\rm v})}_{{\bm k}, \tau, s}$ and eigenvectors $|u^{{\rm c} ({\rm v})}_{{\bm k}, \tau, s}\rangle$ are:
\begin{equation}\label{eq:eigenvalues}
\epsilon^{{\rm c} ({\rm v})}_{{\bm k}, \tau, s} = h_{0}({\bm  k},\tau, s)\pm\sqrt{\left[h_{z}({\bm  k},\tau, s)\right]^2+|h_{12}({\bm  k},\tau, s)|^2}
\end{equation}
and
\begin{equation}\label{eq:eigenspinors}
|u^{{\rm c} ({\rm v})}_{{\bm k}, \tau, s}\rangle =\frac{1}{\sqrt{\left[D^{{\rm c} ({\rm v})}({\bm  k},\tau, s)\right]^2+|h_{12}({\bm  k},\tau, s)|^2}}
\begin{bmatrix}
-h_{12}({\bm  k},\tau, s) \\ D^{{\rm c} ({\rm v})}({\bm  k},\tau, s)
\end{bmatrix}~,
\end{equation}
where
\begin{equation}
h_{0}({\bm  k},\tau, s)=\frac{\lambda_0}{2}\tau s +\frac{\hbar^2 k^2}{4m_0}\alpha+t_2a^3_0\tau(k^3_x-3k_xk^2_y)\alpha'~,
\end{equation}
\begin{equation}
h_{z}({\bm  k},\tau, s) = \frac{\Delta+\lambda \tau s}{2} +\frac{\hbar^2 k^2}{4m_0}\beta + t_2a^3_0\tau(k^3_x-3k_xk^2_y)\beta'~,
\end{equation}
\begin{equation}
h_{12}({\bm  k},\tau, s)= t_0a_0(\tau k_x-i k_y)+t_1 a^2_0(\tau k_x+i k_y)^2~,
\end{equation}
and
\begin{eqnarray}
D^{{\rm c} ({\rm v})}({\bm  k},\tau, s) &=& h_{z}({\bm  k}, \tau, s)
\nonumber\\ &\mp&
\sqrt{\left[h_{z}({\bm  k},\tau, s)\right]^2+|h_{12}({\bm  k},\tau, s)|^2}~.
\end{eqnarray}
\par
We need the matrix elements of the current operator for the evaluation of the nonlinear response functions. We start by introducing the so-called paramagnetic current operator\cite{Giuliani_and_Vignale} ($c= 1$, where $c$ is the speed of light, $-e<0$ is the electron charge):
\begin{equation}
j_{\ell}({\bm k}) \equiv - \left.\frac{\delta {\cal H}({\bm k} + e {\bm A}/\hbar)}{\delta A_{\ell}}\right|_{{\bm A} = {\bm 0}}
= - \frac{e}{\hbar}\frac{\partial {\cal H}}{\partial k_{\ell}}~,
\end{equation}
where $\ell = x,y$ is a Cartesian index. The diamagnetic contributions to the current operator can be written as follows\cite{RKP16}:
\begin{equation}
\kappa_{\ell_1\ell_2}({\bm k}) \equiv - \left.\frac{\delta^2 {\cal H}({\bm k} + e {\bm A}/\hbar)}{\delta A_{\ell_1} \delta  A_{\ell_2}}\right|_{{\bm A} = {\bm 0}}
= - \left ( \frac{e}{\hbar} \right)^2 \frac{\partial^2 {\cal H}}{\partial k_{\ell_1} \partial k_{\ell_2}}
\end{equation}
and
\begin{eqnarray}
\xi_{\ell_1\ell_2\ell_3}({\bm k}) &\equiv& -\left.\frac{\delta^3 {\cal H}({\bm k} + e {\bm A}/\hbar)}{\delta A_{\ell_1} \delta  A_{\ell_2} \delta  A_{\ell_3}}\right|_{{\bm A} = {\bm 0}} \\
\nonumber \\
&=& - \left (\frac{e}{\hbar} \right)^3 \frac{\partial^3 {\cal H}}{\partial k_{\ell_1} \partial k_{\ell_2} \partial k_{\ell_3}}
\end{eqnarray}
Using the continuum-model Hamiltonian introduced in Eqs. (\ref{eq:H_iso}) and (\ref{eq:H_tw}), we find:
\begin{equation}
j_\ell = -\frac{e}{\hbar} \left \{\right. \frac{ \partial h_0}{\partial k_\ell} + \frac{\partial h_z}{\partial k_\ell} \sigma_z + {\rm Re}[\frac{\partial h_{12}}{\partial k_\ell}]\sigma_x
 -{\rm Im}[\frac{\partial h_{12}}{\partial k_\ell}] \sigma_y \left.\right \}
\end{equation}
and
\begin{eqnarray}
\kappa_{\ell\ell} &=& -\left (\frac{e}{\hbar} \right )^2 \Bigg \{ \frac{ \partial^2 h_0}{\partial k^2_\ell} + \frac{\partial^2 h_z}{\partial k^2_\ell} \sigma_z + {\rm Re}[\frac{\partial^2 h_{12}}{\partial k^2_\ell}]\sigma_x
\nonumber \\ &-&
{\rm Im}[\frac{\partial^2 h_{12}}{\partial k^2_\ell}] \sigma_y \Bigg \}~.
\end{eqnarray}
Similarly, one can derive an explicit expression for $\xi_{\ell\ell\ell}$.

The required matrix elements of $j_\ell$ and $\kappa_{\ell\ell}$ between the eigenspinors (\ref{eq:eigenspinors}) are given by:
\begin{widetext}
\begin{eqnarray}\label{eq:j-tw}
j^{\rm cv}_\ell({\bm k},\tau, s) \equiv \langle u^{\rm c}_{{\bm k}, \tau, s}| j_\ell | u^{\rm v}_{{\bm k}, \tau, s}\rangle
&=& \frac{e }{\hbar}\Biggl \{
\frac{h_z({\bm  k},\tau, s)
{\rm Re}\left[h_{12}({\bm  k},\tau, s)  {\partial h^\ast_{12}({\bm  k},\tau, s)}/{\partial k_\ell} \right]}
{
|h_{12}({\bm  k},\tau, s)|\sqrt{\left[h_z({\bm  k},\tau, s)\right]^2+\left|h_{12}({\bm  k},\tau, s)\right|^2}
}
+
i\frac{
{\rm Im}\left[h_{12}({\bm  k},\tau, s)  {\partial h^\ast_{12}({\bm  k},\tau, s)}/{\partial k_\ell} \right]}
{
|h_{12}({\bm  k},\tau, s)|
}
\nonumber \\
&-&
\frac{|h_{12}({\bm  k},\tau, s)|  {\partial h_{z}({\bm  k},\tau, s)}/{\partial k_\ell}}
{
\sqrt{\left[h_z({\bm  k},\tau, s)\right]^2+\left|h_{12}({\bm  k},\tau, s)\right|^2}
}
\Biggr \}~,
\end{eqnarray}
\begin{equation}
j^{{\rm cc} ({\rm vv})}_\ell({\bm  k},\tau, s)  \equiv \langle u^{{\rm c}({\rm v})}_{{\bm k}, \tau, s}| j_\ell | u^{{\rm c}({\rm v})}_{{\bm k}, \tau, s}\rangle =  - \frac{e }{\hbar}\Biggl\{
\frac{\partial h_0({\bm  k},\tau, s)}{\partial k_\ell}
\pm \frac{h_z({\bm  k},\tau, s) \partial h_z({\bm  k},\tau, s)/{\partial k_\ell} + {\rm Re}\left[h_{12}({\bm  k},\tau, s)  {\partial h^\ast_{12}({\bm  k},\tau, s)}/{\partial k_\ell} \right]}{\sqrt{\left[h_z({\bm  k},\tau, s)\right]^2+\left|h_{12}({\bm  k},\tau, s)\right|^2}}
\Biggr\}~,
\end{equation}
\begin{eqnarray}\label{eq:j-tw}
\kappa^{\rm cv}_{\ell\ell}({\bm  k},\tau, s) &\equiv& \langle u^{\rm c}_{{\bm k}, \tau, s}| \kappa_{\ell \ell} | u^{\rm v}_{{\bm k}, \tau, s}\rangle
= \left ( \frac{e }{\hbar}\right)^2 \Biggl \{
\frac{h_z({\bm  k},\tau, s)
{\rm Re}\left[h_{12}({\bm  k},\tau, s)  {\partial^2 h^\ast_{12}({\bm  k},\tau, s)}/{\partial k^2_\ell} \right]}
{
|h_{12}({\bm  k},\tau, s)|\sqrt{\left[h_z({\bm  k},\tau, s)\right]^2+\left|h_{12}({\bm  k},\tau, s)\right|^2}
}
\nonumber\\
&+&
i\frac{
{\rm Im}\left[h_{12}({\bm  k},\tau, s)  {\partial^2 h^\ast_{12}({\bm  k},\tau, s)}/{\partial k^2_\ell} \right]}
{
|h_{12}({\bm  k},\tau, s)|
}
-
\frac{|h_{12}({\bm  k},\tau, s)|  {\partial^2 h_{z}({\bm  k},\tau, s)}/{\partial k^2_\ell}}
{
\sqrt{\left[h_z({\bm  k},\tau, s)\right]^2+\left|h_{12}({\bm  k},\tau, s)\right|^2}
}
\Biggr \}~,
\end{eqnarray}
and
\begin{eqnarray}
\kappa^{{\rm cc} ({\rm vv})}_{\ell\ell}({\bm  k},\tau, s)  &\equiv& \langle u^{{\rm c}({\rm v})}_{{\bm k}, \tau, s}| \kappa_{\ell\ell} | u^{{\rm c}({\rm v})}_{{\bm k}, \tau, s}\rangle
=  - \left (\frac{e }{\hbar} \right )^2 \Biggl\{
\frac{\partial^2 h_0({\bm  k},\tau, s)}{\partial k^2_\ell}
\nonumber\\
&\pm& \frac{h_z({\bm  k},\tau, s) \partial^2 h_z({\bm  k},\tau, s)/{\partial k^2_\ell} + {\rm Re}\left[h_{12}({\bm  k},\tau, s)  {\partial^2 h^\ast_{12}({\bm  k},\tau, s)}/{\partial k^2_\ell} \right]}{\sqrt{\left[h_z({\bm  k},\tau, s)\right]^2+\left|h_{12}({\bm  k},\tau, s)\right|^2}}
\Biggr\}~.\nonumber \\
\end{eqnarray}
\end{widetext}
We note that intra-band matrix elements (e.g. $j^{\rm cc}_y$ and $\kappa^{\rm cc}_{yy}$) have a definite parity while inter-band ones (e.g. $j^{\rm cv}_y$ and $\kappa^{\rm cv}_{yy}$) do not. This is at the origin of the vanishing of the paramagnetic contribution to even harmonic-generation response functions. Therefore, as we will see later, only diamagnetic terms yield a finite contribution to even harmonic-generation responses.
\subsection{General symmetry considerations}
\begin{figure}[h!]
\centerline{\includegraphics[width=80mm]{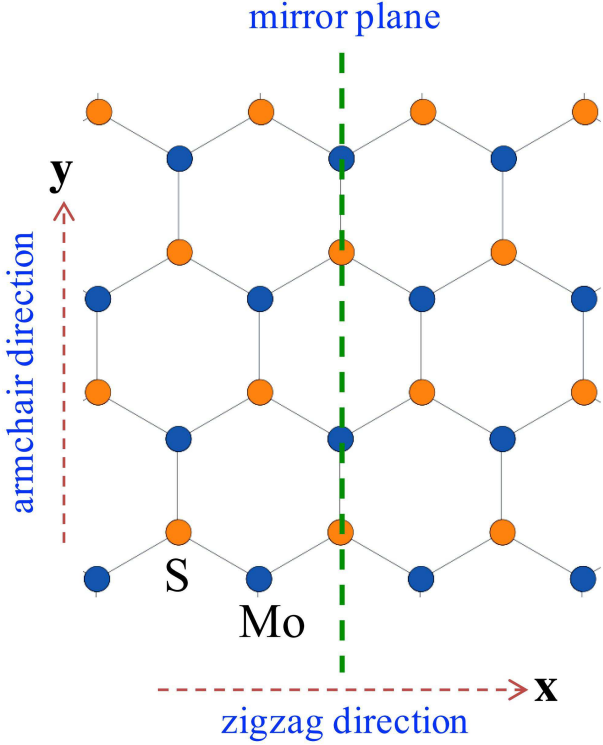}}
\caption{Top view of the 1L-MoS$_2$ lattice.\label{fig:lattice}}
\end{figure}
Our continuum-model Hamiltonian is derived from a tight-binding Hamiltonian in which the zigzag direction of the lattice coincides with the ${\hat {\bm x}}$ direction. The zigzag direction is perpendicular to the reflection (mirror) symmetry plane of the 1L-MoS$_2$ lattice (see Fig.\ref{fig:lattice}).

The $n$-th order optical susceptibilities $\chi^{(n)}_{\ell i_1i_2\dots i_n}$ are defined as:
\begin{eqnarray}\label{eq:perturb2}
P^{(n)}_{\ell}(\omega_\Sigma) &=& \epsilon_0 \sum_{i_1 i_2\dots i_n}  \chi^{(n)}_{\ell i_1 i_2\dots i_n }(-\omega_\Sigma;\omega_1,\omega_2,\dots,\omega_n)
\nonumber \\ &\times&
E_{i_1}(\omega_1)E_{i_2}(\omega_2)\dots E_{i_n}(\omega_n)~,
\end{eqnarray}
where $E_i$ and ${P}^{(n)}_{\ell}$ are the Cartesian components of the electric field ${\bm E}$ and the $n$-th order macroscopic polarization ${\bm  P}^{(n)}$, respectively, and $\epsilon_0$ is the vacuum permittivity. Note that $i_1$, $i_2$, \dots, $i_n$ are Cartesian indices and $\omega_\Sigma \equiv \sum_i\omega_i$.

Since 1L-MoS$_2$ belongs to the $D^1_{3h}$ symmetry group, the only non-vanishing elements of the second-order susceptibility are\cite{boyd}:
\begin{equation}\label{eq:sym2}
\chi^{(2)}_{yyy}=-\chi^{(2)}_{yxx}=-\chi^{(2)}_{xxy}=-\chi^{(2)}_{xyx}~,
\end{equation}
while for the case of the third-order response we have\cite{boyd}:
\begin{equation}\label{eq:sym3_part1}
\chi^{(3)}_{yyyy}=\chi^{(3)}_{xxxx}=\chi^{(3)}_{yyxx}+\chi^{(3)}_{yxxy}+\chi^{(3)}_{yxyx}~,
\end{equation}
and
\begin{eqnarray}\label{eq:sym3_part2}
&&\chi^{(3)}_{xxyy}=\chi^{(3)}_{yyxx}~,\nonumber\\
&&\chi^{(3)}_{xyyx}=\chi^{(3)}_{yxxy}~,\nonumber\\
&&\chi^{(3)}_{xyxy}=\chi^{(3)}_{yxyx}~.
\end{eqnarray}
In the case of a linearly-polarized pump laser, we expect a SHG maximum when the laser is polarized along the $\hat{\bm y}$ direction, i.e. perpendicular to the zigzag direction. On the contrary, if the incident light is polarized along the $\hat{\bm x}$ direction, i.e. the zigzag direction, we expect a vanishing SHG signal due to the reflection symmetry (i.e.~$\sigma_{\rm v}:x\to-x$) along this axis. Our continuum-model Hamiltonian is consistent with these general expectations based on symmetry. We find $\chi^{(2)}_{xxx}=0$, even in the presence of trigonal warping, because the contribution in the two valleys identically cancel each other.

Using Eqs.\ref{eq:perturb2},\ref{eq:sym2},\ref{eq:sym3_part1},\ref{eq:sym3_part2} we obtain Eqs.3,4 of the main text, which describe the dependence between induced charge polarization, ${\bm P}$, and the polarization of the incident laser. In the case of a circularly-polarized pump laser, we have ${\bm E}=|{\bm E}|{\hat {\bm \varepsilon}}_{\pm}$ with ${\hat {\bm \varepsilon}}_{\pm} = (\hat {\bm x} \pm i \hat{\bm y})/\sqrt{2}$. Using Eqs.3,4 of the main text we arrive at the following results for the circularly-polarized pump laser:
\begin{equation}\label{eq:secondorder-circ2}
{\bm P}^{(2)}= \mp i\sqrt{2} \epsilon_0 \chi^{(2)}_{yyy} |\bm E|^2  \hat{\bm \varepsilon}_{\mp}
\end{equation}
and
\begin{equation}\label{eq:thirdorder-circ2}
{\bm P}^{(3)}=0~.
\end{equation}
Eq.\ref{eq:secondorder-circ2} implies an {\it opposite} polarization of the SHG signal with respect to the laser, while Eq.\ref{eq:thirdorder-circ2} implies no THG signal in response to a circularly-polarized pump laser.

For quantitative results, only the three tensor elements $\chi^{(2)}_{yyy}$, $\chi^{(3)}_{yyyy}$ and $\chi^{(4)}_{yyyy}$ are required for second-, third- and fourth-order nonlinear response functions.
\subsection{Nonlinear response functions}
The response of an electron system to light can be calculated by adopting different gauges for describing the electric field of incident light. The gauge in which a uniform electric field ${\bm E}(t)$ is described in terms of a uniform time-dependent vector potential, ${\bm E}(t) = -\partial {\bm A}(t)/\partial t$, is convenient in solids as it does not break Bloch translational invariance. The vector potential couples to matter degrees of freedom through the minimal coupling, i.e.~${\bm k} \to {\bm k} + e {\bm A}/\hbar$. The external vector potential induces a current ${\bm J}(t)$, which can be expanded in a power series of ${\bm A}(t)$. For each Cartesian component, $J_\ell = \sum_n J^{(n)}_{\ell}$ where $n$ denotes the $n$-th order in powers of ${\bm A}(t)$. In Fourier transform with respect to time we get:
\begin{eqnarray}\label{eq:perturb}
J^{(n)}_{\ell}(\omega_\Sigma) &\equiv& \sum_{i_1, i_2, \dots, i_n}\Pi^{(n)}_{\ell i_1 i_2\dots i_n}(-\omega_\Sigma;\omega_1,\omega_2,\dots,\omega_n)
\nonumber \\ &\times& A_{i_1}(\omega_1)A_{i_2}(\omega_2)\dots A_{i_n}(\omega_n)~,
\end{eqnarray}
where ${\bm  A}(\omega_i)=-i {\bm  E}(\omega_i)/(\omega_i+i\eta/\hbar)$ and $\eta$ is an infinitesimal positive real number, needed to ensure that the external field is absent in the remote past ($t \rightarrow -\infty$).

Since the macroscopic current is related to the macroscopic polarization by
${\bm  J}(t) = \partial {\bm P}/\partial t$~\cite{Griffiths}, we get ${\bm  J}^{(n)}(\omega_\Sigma) = - i(\omega_\Sigma+i\eta/\hbar) {\bm  P}^{(n)}(\omega_\Sigma)$, for each order in perturbation theory.

We finally find the following relation between nonlinear response functions and optical susceptibilities:
\begin{eqnarray}
&&\epsilon_0\chi^{(n)}_{\ell i_1 i_2 \dots i_n }(-\omega_\Sigma;\omega_1,\dots,\omega_n) =  i(-i)^n \times
\nonumber \\
&&  \frac{\Pi^{(n)}_{\ell i_1 i_2 \dots i_n }(-\omega_\Sigma;\omega_1,\dots,\omega_n)}{(\omega_\Sigma+i\eta/\hbar)(\omega_n+i\eta/\hbar)\dots(\omega_1+i\eta/\hbar)}~.
\end{eqnarray}
The $n$-th order nonlinear response $\Pi^{(n)}_{\ell i_1 i_2\dots i_n }$ contains both paramagnetic and diamagnetic current contributions, which will be denoted by $\Pi^{(n),{\rm P}}_{\ell i_1 i_2\dots i_n }$ and $\Pi^{(n),{\rm D}}_{\ell i_1 i_2\dots i_n }$, respectively. The  paramagnetic current correlators, which are diagrammatically illustrated in Fig.~\ref{fig:paramagnetic}, read:
\begin{figure*}
\centerline{\includegraphics[width=170mm]{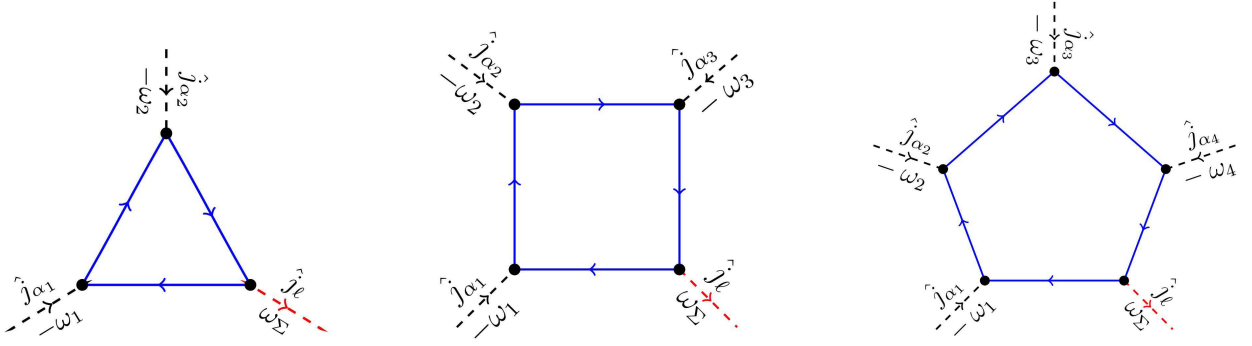}}
\caption{Three-, four-, and five-leg Feynman diagrams for the second-, third-, and fourth-order nonlinear {\em paramagnetic} response functions. Solid lines denote electron propagators while dashed lines denote photons. The quantities $\omega_1=\dots=\omega_4=\omega$ indicate the incoming photon frequencies, while $\hat{j}_\alpha$ denotes the $\alpha$-th Cartesian component of the paramagnetic current operator.}
\label{fig:paramagnetic}
\end{figure*}
\begin{equation}\label{eq:first}
\Pi^{(1),{\rm P}}_{\ell i_1 }(i\nu) \equiv \left \langle \hat j_{i_1}(-i\nu) \hat j_{\ell}(i\nu) \right \rangle~,
\end{equation}
\begin{equation}\label{eq:second}
\Pi^{(2),{\rm P}}_{\ell i_1 i_2 }(-i\nu_{\Sigma};i\nu_1 ,i\nu_2) \equiv \sum'_{\cal P}
\left \langle \hat j_{i_1}(-i\nu_1)\hat j_{i_2}(-i\nu_2 ) \hat j_{\ell}(i\nu_{\Sigma}) \right \rangle~,
\end{equation}
\begin{eqnarray}\label{eq:third}
&&\Pi^{(3),{\rm P}}_{\ell i_1 i_2 i_3 }(-i\nu_{\Sigma};i\nu_1 ,i\nu_2 ,i\nu_3)
\nonumber \\ & \equiv& \sum'_{\cal P}
\left \langle \hat j_{i_1}(-i\nu_1)\hat  j_{i_2}(-i\nu_2) \hat j_{i_3}(-i\nu_3 ) \hat j_{\ell}(i\nu_{\Sigma}) \right \rangle~,
\end{eqnarray}
and
\begin{eqnarray}\label{eq:fourth}
&&\Pi^{(4),{\rm P}}_{\ell i_1 i_2 i_3 i_4 }(-i\nu_{\Sigma};i\nu_1,i\nu_2 ,i\nu_3 ,i\nu_4) \nonumber\\
&\equiv&\sum'_{\cal P}
\left\langle \hat j_{i_1}(-i\nu_1) \hat j_{i_2}(-i\nu_2)  \hat j_{i_3}(-i\nu_3 ) \hat j_{i_4}(-i\nu_4) \hat j_{\ell}(i\nu_{\Sigma}) \right \rangle~.
\nonumber\\
\end{eqnarray}
\begin{figure*}
\centerline{\includegraphics[width=120mm]{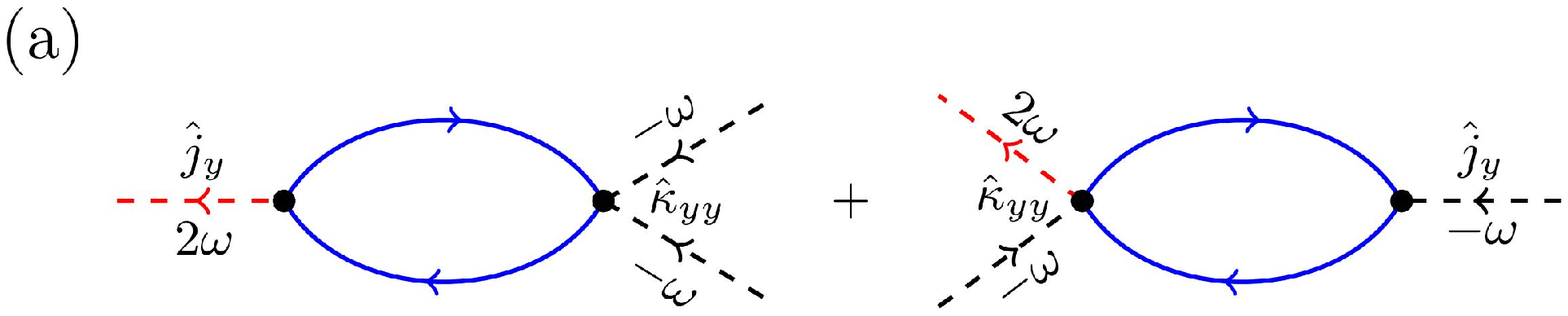}}
\centerline{\includegraphics[width=160mm]{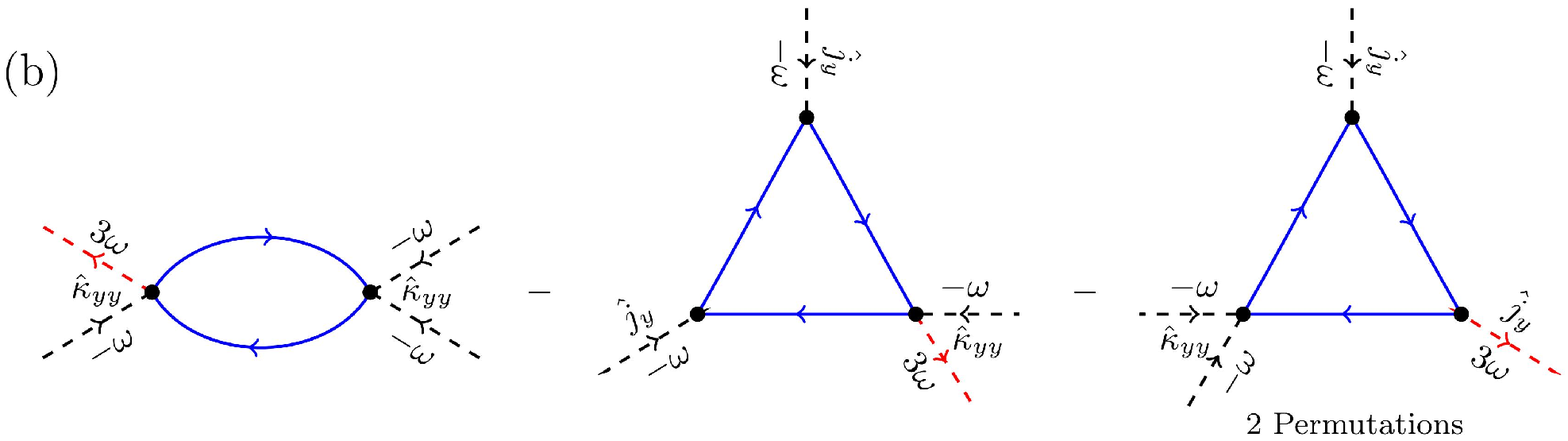}}
\caption{Feynman diagrams for the {\it diamagnetic} contributions to the second- and third-order response functions. a) second-order response. b) third-order response.}
\label{fig:diamagnetic}
\end{figure*}

Here, $\langle\dots\rangle$ denotes the thermal averaging\cite{Giuliani_and_Vignale,Mahan}, $\hat j_{i}$ indicates the second-quantized form of $i$-th Cartesian component of the paramagnetic current operator, $\sum'_{\cal P}$ enforces the so-called ``{\em intrinsic permutation symmetry}'' among all dummy variables $(i_n,\nu_n)$\cite{Butcher_and_Cotter}, and  $\nu_{\Sigma}=\sum_i \nu_i$, where $\nu_i=2\pi n/\beta$'s are bosonic Matsubara energies corresponding to the incident photon energies. $n$ is a relative integer. $\beta=1/(k_{\rm B}T)$, with T the electron temperature.

The paramagnetic current correlators in Eqs.\ref{eq:first}-\ref{eq:fourth} can be calculated by using the many-body diagrammatic perturbation theory\cite{J12,rostami_prb_2016}. Following Ref.\cite{rostami_prb_2016}, we first sum over the Fermionic Matsubara energies and then carry out the analytical continuation $\nu_i=\nu \to \hbar\omega+i\eta$ where $\eta \to 0^+$. We find the following relations for the case of $\ell=i_n =y$:
\begin{equation}\label{eq:pi1}
\Pi^{(1),{\rm P}}_{yy}(\omega)=  \sum_{{\bm k}, \tau, s}\sum_{\{\lambda_i\}}U_{\lambda_1\lambda_2}
j^{\lambda_2\lambda_1}_{y} j^{\lambda_1\lambda_2}_{y}~,
\end{equation}
\begin{eqnarray}\label{eq:pi2}
\Pi^{(2),{\rm P}}_{yyy}(-2\omega;\omega,\omega)
&=& \sum_{{\bm k}, \tau, s}\sum_{\{\lambda_i\}} \frac{ j^{\lambda_3\lambda_2}_{y}  j^{\lambda_2\lambda_1}_{y}  j^{\lambda_1\lambda_3}_{y}}{2(\hbar\omega+i \eta) + \epsilon^{\lambda_1}_{{\bm k}, \tau, s} - \epsilon^{\lambda_3}_{{\bm k}, \tau, s}}
\nonumber \\ &\times&
\left(U_{\lambda_1\lambda_2} - U_{\lambda_2\lambda_3}\right)~,
\end{eqnarray}
\begin{widetext}
\begin{equation}\label{eq:pi3}
\Pi^{(3),{\rm P}}_{yyyy}(-3\omega;\omega,\omega,\omega)=
\sum_{{\bm k},\tau,s}\sum_{\{\lambda_i\}}
\frac{j^{\lambda_4\lambda_3}_{y} j^{\lambda_3\lambda_2}_{y}  j^{\lambda_2\lambda_1}_{y}  j^{\lambda_1\lambda_4}_{y}}{3(\hbar\omega+i\eta) +\epsilon^{\lambda_1}_k-\epsilon^{\lambda_4}_k}
 \Biggl\{\frac{U_{\lambda_1\lambda_2} - U_{\lambda_2\lambda_3}}{2(\hbar\omega +i \eta) +\epsilon^{\lambda_1}_{{\bm k},\tau,s}-\epsilon^{\lambda_3}_{{\bm k},\tau,s}}
 -\frac{U_{\lambda_2\lambda_3}-U_{\lambda_3\lambda_4}}{2(\hbar\omega + i\eta) + \epsilon^{\lambda_2}_{{\bm k},\tau,s} - \epsilon^{\lambda_4}_{{\bm k},\tau,s}}\Biggr\}~,
\end{equation}
and
\begin{eqnarray}\label{eq:pi4}
&&\Pi^{(4),{\rm P}}_{yyyyy}(-4\omega;\omega,\omega,\omega,\omega)=
 \sum_{{\bm k}, \tau, s} \sum_{\{\lambda_i\}}
 \frac{ j^{\lambda_5\lambda_4}_{y} j^{\lambda_4\lambda_3}_{y} j^{\lambda_3\lambda_2}_{y} j^{\lambda_2\lambda_1}_{y}  j^{\lambda_1\lambda_5}_{y}}
  {4(\hbar \omega+i\eta)+\epsilon^{\lambda_1}_{{\bm k},\tau,s} - \epsilon^{\lambda_5}_{{\bm k},\tau,s}}
\nonumber\\&&
\Bigg\{
\frac{1}{3(\hbar\omega + i \eta) + \epsilon^{\lambda_1}_{{\bm k},\tau,s} - \epsilon^{\lambda_4}_{{\bm k},\tau,s}}
\Bigg[\frac{U_{\lambda_1\lambda_2}-U_{\lambda_2\lambda_3}}{2(\hbar\omega+i\eta) +\epsilon^{\lambda_1}_{{\bm k},\tau,s} - \epsilon^{\lambda_3}_{{\bm k},\tau,s} } -\frac{U_{\lambda_2\lambda_3}-U_{\lambda_3\lambda_4}}{2(\hbar\omega + i\eta) +\epsilon^{\lambda_2}_{{\bm k},\tau,s} - \epsilon^{\lambda_4}_{{\bm k},\tau_s}} \Bigg]
 \nonumber\\ &&
-\frac{1}{3(\hbar\omega+i\eta)+\epsilon^{\lambda_2}_{{\bm k},\tau,s} - \epsilon^{\lambda_5}_{{\bm k},\tau,s}}
\Bigg[\frac{U_{\lambda_2\lambda_3} - U_{\lambda_3\lambda_4} }{2(\hbar\omega+i\eta) + \epsilon^{\lambda_2}_{{\bm k},\tau,s} - \epsilon^{\lambda_4}_{{\bm k},\tau,s}}-\frac{U_{\lambda_3\lambda_4} - U_{\lambda_4\lambda_5}}{2(\hbar\omega+i\eta) +\epsilon^{\lambda_3}_{{\bm k},\tau,s} - \epsilon^{\lambda_5}_{{\bm k},\tau,s}}
 \Bigg]
 \Bigg\}~.
\end{eqnarray}
\end{widetext}
For simplicity, we introduce the quantity $U_{\lambda \lambda'}$ as follows:
\begin{equation}
U_{\lambda\lambda'}({\bm k},\omega, \tau, s) \equiv \frac{1}{\cal S} \frac{n_{\rm F}(\epsilon^\lambda_{{\bm k}, \tau, s}) - n_{\rm F}(\epsilon^{\lambda'}_{{\bm k}, \tau, s})}{\hbar \omega +\epsilon^{\lambda}_{{\bm k}, \tau, s}-\epsilon^{\lambda'}_{{\bm k}, \tau, s}+i\eta}~,
\end{equation}
where ${\cal S}$ is the sample area, $\lambda, \lambda' = {\rm c}, {\rm v}$, and
\begin{equation}
n_{\rm F}(E) = \left\{\exp\left(\frac{E-\mu}{k_{\rm B} T}\right)+1\right\}^{-1}
\end{equation}
is the Fermi-Dirac distribution function at finite temperature $T$ and chemical potential $\mu$. In Eqs.\ref{eq:pi1}-\ref{eq:pi4} we dropped the explicit functional dependence on ${\bm k},\tau,s$, e.g. $j^{mn}_{y} = j^{mn}_{y}({\bm k}, \tau, s)$ and $U_{mn} = U_{mn}({\bm k},\omega, \tau, s)$.  We find most convenient to first carry out the sum over the band indices $\lambda_i$ and then carry out numerically the integral over the wave vector ${\bm k}$.

The paramagnetic contributions to the even-order response functions, $\Pi^{(2),{\rm P}}_{yyy}$ and $\Pi^{(4),{\rm P}}_{yyyyy}$, vanish because $\epsilon^{{\rm c} ({\rm v})}_{{\bm k}, \tau, s}$ is an even function of $k_y$. This property of the energy dispersion is protected by symmetry, and stems from time-reversal (${\cal T}$) and reflection ($\sigma_{\rm v}$) symmetries.

A microscopic calculation of even-order response functions requires the knowledge of diamagnetic contributions. These can be included with the aid of correlation functions involving the $\hat \kappa_{yy}$ operator. In fact, $\hat \xi_{yyy}$ could also contribute to diamagnetic responses. However, in our low-energy model $\hat \xi_{yyy}$ is identically zero. Similar to the paramagnetic case, $\hat \kappa_{yy}$ and $\hat \xi_{yyy}$ indicate the second-quantized form of the diamagnetic current operators (i.e. $\kappa_{yy}$ and $\xi_{yyy}$). Diamagnetic contributions to the second- and third-order response functions are reported in Fig.\ref{fig:diamagnetic}, in terms of Feynman diagrams. For the sake of simplicity, we have not calculated diamagnetic contributions to the fourth-order response.

According to Fig.\ref{fig:diamagnetic}a, the diamagnetic contribution to the second-order response is given by:
\begin{eqnarray}\label{eq:pi2_dia}
\Pi^{(2),{\rm D}}_{yyy}(-2\omega;\omega,\omega) = & -& \sum_{{\bm k}, \tau, s}\sum_{\{\lambda_i\}} \Bigg [
U_{\lambda_1\lambda_2}
j^{\lambda_1\lambda_2}_{y}  \kappa^{\lambda_2\lambda_1}_{yy}
\nonumber \\
&+&\widetilde U_{\lambda_1\lambda_2}
\kappa^{\lambda_1\lambda_2}_{yy} j^{\lambda_2\lambda_1}_{y} \Bigg ]~.
\end{eqnarray}
Similarly, the diamagnetic contribution to the third-order response,Fig.\ref{fig:diamagnetic}b,is given by:
\begin{eqnarray}\label{eq:pi3}
&&\Pi^{(3),{\rm D}}_{yyyy}(-3\omega;\omega,\omega,\omega)
= \sum_{{\bm k}, \tau, s}\sum_{\{\lambda_i\}} \Bigg \{
\widetilde U_{\lambda_1\lambda_2}
\kappa^{\lambda_1\lambda_2}_{yy}  \kappa^{\lambda_2\lambda_1}_{yy}
\nonumber \\&-&
 \frac{ j^{\lambda_3\lambda_2}_{y}  j^{\lambda_2\lambda_1}_{y}  \kappa^{\lambda_1\lambda_3}_{yy}}{2(\hbar\omega+i \eta) + \epsilon^{\lambda_1}_{{\bm k}, \tau, s} - \epsilon^{\lambda_3}_{{\bm k}, \tau, s}} \left(U_{\lambda_1\lambda_2} - U_{\lambda_2\lambda_3}\right)
\nonumber \\ &-&
\sum'_{\cal P} \frac{ \kappa^{\lambda_3\lambda_2}_{yy}  j^{\lambda_2\lambda_1}_{y}  j^{\lambda_1\lambda_3}_{y}}{3(\hbar\omega+i \eta) + \epsilon^{\lambda_1}_{{\bm k}, \tau, s} - \epsilon^{\lambda_3}_{{\bm k}, \tau, s}} \left(\widetilde U_{\lambda_1\lambda_2}  - U_{\lambda_2\lambda_3}\right)~.\Bigg \}
\nonumber \\
\end{eqnarray}
Here, $\widetilde U_{\lambda_1\lambda_2}= U_{\lambda_1\lambda_2} ({\bm k},2\omega,\tau,s)$ with $\kappa^{mn}_{yy} = \kappa^{mn}_{yy}({\bm k}, \tau, s)$ is the matrix element of $\kappa_{yy}$.

Since our low-energy model is valid for a limited range of values of the wave vector ${\bm k}$, we must introduce an ultra-violet cut-off, which breaks gauge invariance\cite{chirolli_prl_2012}. We therefore need to regularize our final results to avoid unphysical response function. This can be accomplished\cite{chirolli_prl_2012} by considering the following gauge-regularized response tensors: $\Pi^{(n)}_{\ell i_1 i_2\dots i_n } \equiv \Pi^{(n)}_{\ell i_1 i_2\dots i_n } -\Pi^{(n)}_{\ell i_1 i_2\dots i_n }\big |_{\{\omega_i\} \to 0}$.

We note that the summands in Eqs.\ref{eq:pi2},\ref{eq:pi4},\ref{eq:pi2_dia} contain an {\it odd} number of matrix elements of the paramagnetic ($j_y$) and diamagnetic ($\kappa_{yy}$) current operators. In the absence of trigonal warping, the overall form-factor, which is proportional to these matrix elements, is an odd function of $k_y$: we therefore conclude that, {\it in the absence of trigonal warping}, $\Pi^{(2)}_{yyy}(-2\omega;\omega,\omega) = \Pi^{(4)}_{yyyyy}(-4\omega;\omega,\omega,\omega,\omega) =0$. An identical conclusion was reached for other isotropic low-energy continuum model Hamiltonians, such as those describing gapped graphene\cite{MMG13} and biased bilayer graphene\cite{WX12,BP15}. We expect the second-order nonlinear response function $\Pi^{(2)}_{yyy}$ to be small compared to the third-order one, since it is controlled by a small trigonal warping correction (${\cal H}_{\rm tw}$) in comparison with the fully isotropic leading term (${\cal H}_{\rm i}$) in the low-energy model Hamiltonian. Of course, this conclusion is valid within the single-particle picture and in the low-energy limit, which we have relied on so far.
\subsection{Relative magnitude of nonlinear responses: ratios of irradiances}
To quantify the {\it relative} magnitude of nonlinear harmonic signals, we calculate ratios between induced polarizations $P^{(n)}_y$ at different orders $n$ in perturbation theory. For a linearly-polarized laser (e.g.~${\bm E}=|{\bm E}| {\hat {\bm y}}$):
\begin{widetext}
\begin{eqnarray}\label{eq:ratio}
\left|\frac{P^{(n+1)}_y}{P^{(n)}_y}\right|&=&\left|\frac{\chi^{(n+1)}_{\underbrace{y\ldots y}_{n+2\ \text{times}}}|{\bm  E}|}{\chi^{(n)}_{\underbrace{y\ldots y}_{n+1\ \text{times}}} }\right|=
\left|\frac{\Pi^{(n+1)}_{\underbrace{y\ldots y}_{n+2\ \text{times}}}/\Pi^{(n+1)}_0}{(\hbar\omega+i\eta)/({\rm eV})\times \Pi^{(n)}_{\underbrace{y\ldots y}_{n+1\ \text{times}}}/\Pi^{(n)}_0}\right|
\times
\left (\frac{ n \Pi^{(n+1)}_0\hbar}{(n+1) \Pi^{(n)}_0({\rm eV})}\right ) \times |{\bm  E}|
\nonumber\\
&=&\frac{n}{n+1} \times  \frac{t_0}{{\rm eV}}\times\frac{a_0}{\rm m} \times\frac{|{\bm  E}|}{{\rm V}/{\rm m}}\times X_{n+1,n}(\omega)~,
\end{eqnarray}
\end{widetext}
where
\begin{eqnarray}
\Pi^{(n)}_0  \equiv \frac{(e t_0 a_0/\hbar)^{n+1}}{8\pi a^2_0 ({\rm eV})^{n}}
&=&\frac{({\rm eV}){\rm m}^{n-1}}{8\pi} \left(\frac{t_0}{\rm eV}\right)^{n+1}
\nonumber \\ &\times&
 \left(\frac{a_0}{\rm m}\right)^{n-1} \left (\frac{e}{\hbar}\right)^{n+1}
\end{eqnarray}
and the quantities $t_0$ and $a_0$ have been introduced in the Hamiltonian ${\cal H}$. Leaving aside pre-factors, $\Pi_0$ represents the physical dimensions of the nonlinear current correlator $\Pi^{(n)}_{\ell i_1 i_2\dots i_n }(-\omega_\Sigma;\omega_1,\omega_2,\dots,\omega_n)$. In the SI system, the unit of $\Pi^{(n)}_0$ is ${\rm C}\hspace{0.5mm}{\rm m}^{n-1}  {\rm V}^{-n}   {\rm s}^{-(n+1)}$. The dimensionless quantities $X_{n+1,n}$ are given by:
\begin{equation}
X_{n+1,n}(\omega)=\left|\frac{\Pi^{(n+1)}_{\underbrace{y\ldots y}_{n+2\ \text{times}}} /\Pi^{(n+1)}_0}{(\hbar\omega+i\eta)/({\rm eV})\times \Pi^{(n)}_{\underbrace{y\ldots y}_{n+1\ \text{times}}} /\Pi^{(n)}_0}\right|~.
\end{equation}
The amplitude of the electric field ($|{\bm E}|$) in Eq.\ref{eq:ratio} can be replaced by the power of the pump laser ($P_{\rm pump}$) by using the following relation:
\begin{equation}\label{eq:pump}
\frac{P_{\rm pump}}{\pi (D/2)^2}=\frac{1}{2} n_r c\epsilon_0  |{\bm  E}|^2~,
\end{equation}
where $D\approx 1.85~{\rm \mu m}$ is the experimental spot size diameter, $n_r\approx1$ is the refractive index of air, $c\approx3\times 10^8~{\rm m}/{\rm s}$ is the speed of light in vacuum, and $\epsilon_0\approx 8.85\times 10^{-12}~{\rm C}/({\rm V}{\rm m})$ is the vacuum electrical permittivity. Using Maxwell's equations, we can obtain the following wave equation in a nonlinear medium\cite{boyd}:
\begin{equation}\label{eq:waveEQ}
\nabla^2 {\bm E}^{(n)} +\left (\frac{\omega_n}{c}\right )^2   {\bm\epsilon}^{(1)}(\omega_n) \cdot {\bm E}^{(n)}  =
 - \frac{1}{\epsilon_0} \left (\frac{\omega_n}{c}\right )^2  {\bm P}^{(n)}~.
\end{equation}
where $n=2,3,\dots$ indicates the order of nonlinearity, ${\bm \epsilon}^{(1)}$ is the linear dielectric tensor and $ {\bm P}^{(n)}$ is the $n$-th order polarization vector. The intensity $I^{(n)}$ of the $n$-th order nonlinear signal is proportional to the square of the induced electric field amplitude $E^{(n)} \propto \omega^2_n P^{(n)}_y$ where $\omega_n=n\omega$ for the harmonic generation case. Replacing Eq.\ref{eq:pump} in Eq.\ref{eq:ratio} we find:
\begin{equation}\label{eq:theory}
 \frac{I^{(n+1)}}{I^{(n)}} = \left (\frac{n+1}{n}\right )^2 \left|\frac{P^{(n+1)}_y}{P^{(n)}_y}\right|^2=R_{n+1,n}(\omega) P_{\rm pump}~,
\end{equation}
where $R_{n+1,n}(\omega)$ in units of $1/{\rm W}$ is given by:
\begin{eqnarray}\label{eq:R}
R_{n+1,n}(\omega) &=& \frac{8[{\rm m}/{\rm s}] [{\rm C}/({\rm V}{\rm m})] }{\pi n_r c \epsilon_0}
\left[\frac{t_0/({\rm eV}) \times a_0/{\rm m}}{D/{\rm m}}\right]^2
\nonumber \\ &\times&
 \left[X_{n+1,n}(\omega)\right]^2 ~.
\end{eqnarray}
If we assume that the spot size of different harmonic-generated signals on the detector are equal to each other, we can write the following relation between power and intensity ratios:
\begin{equation}\label{eq:exp}
\frac{I^{(n+1)}}{I^{(n)}} \approx  \frac{ P_{(n+1)\omega}}{P_{n\omega}}~,
\end{equation}
where $P_{n\omega}$ denotes the signal power of the $n$-th harmonic-generated signal.

\begin{figure}
\centerline{\includegraphics[width=80mm]{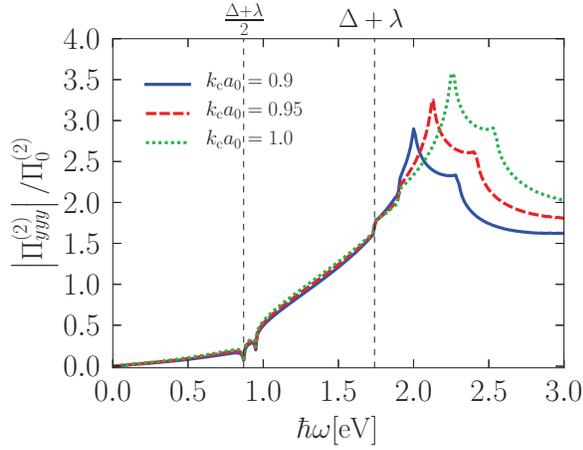}}
\caption{Frequency dependence of the second-order response function $\Pi^{(2)}_{yyy}$ (in units of $\Pi^{(2)}_0$). Different curves refer to different values of the parameter $k_{\rm c}$.\label{fig:SHG}}
\end{figure}

\begin{figure}
\centerline{\includegraphics[width=80mm]{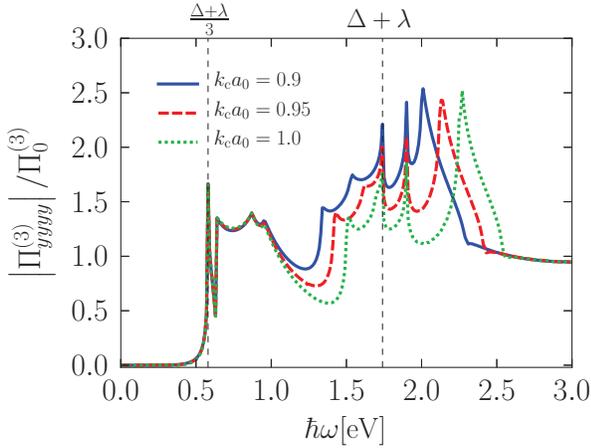}}
\caption{Same as in Fig.\ref{fig:SHG}, but for the case of the third-order response function.\label{fig:THG}}
\end{figure}

\begin{figure}
\centerline{\includegraphics[width=80mm]{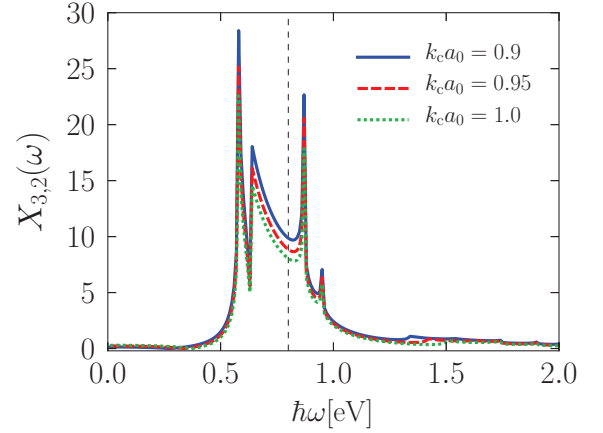}}
\caption{Results for the $X_{3,2}$ as function of the pump laser frequency. Vertical dashed lines is positioned at $\hbar\omega=0.8~{\rm eV}$.\label{fig:X}}
\end{figure}

Our main results for nonlinear response functions of 1L-${\rm MoS}_2$ are summarized in Figs.\ref{fig:SHG}-\ref{fig:X}. We use the following values for the parameters of the model: $\Delta=1.82~{\rm eV}$, $\lambda_0=69~{\rm meV}$, $\lambda=-80~{\rm meV}$,  $t_0=2.34~{\rm eV}$, $\alpha=-0.01$, $\beta=-1.54$, $t_1=-0.14~{\rm eV}$, $t_2=1~{\rm eV}$, $\alpha'=0.44$, and $\beta'=-0.53$. These parameters are obtained from a tight-binding fitting\cite{RG15} of LDA-DFT band structure calculations\cite{CG13,RO14}. In all our numerical results, we use $T=300~{\rm K}$ and $\mu=0$. In Figs.\ref{fig:SHG}-\ref{fig:X}, we check the dependence of our results on the value of the ultra-violet cut-off, $k_{\rm c} \propto 1/a_{0}$. Note that $a_0=a/\sqrt{3}$ with $a\approx 3.16~{\text \AA}$ is the lattice constant of 1L-MoS$_2$.
According to Figs.\ref{fig:SHG},\ref{fig:THG}, the nonlinear response functions start to grow when $\hbar\omega$ is larger than $ (\Delta+\lambda)/2$ and $ (\Delta+\lambda)/3$ for the SHG and THG cases, respectively. $\Delta+\lambda$ is the optical band gap of MoS$_2$. Moreover, in the frequency range of our interest ($< 1~{\rm eV}$) the spectra of the second and third order response functions are not very sensitive to the value of $k_{\rm c}$. The theoretical result shown in Fig.4c) of the main text is obtained by using Eqs.\ref{eq:R},\ref{eq:theory} for $\hbar\omega=0.8~{\rm eV}$.
\section{Acknowledgments}
We thank M. J. Huttunen and R. W. Boyd for useful discussions. We acknowledge funding from the Academy of Finland (No.:276376, 284548), TEKES (NP-Nano,OPEC), Fondazione Istituto Italiano di Tecnologia, the Graphene Flagship, ERC grant Hetero2D, Nokia Foundation, EPSRC grants EP/K01711X/1, EP/K017144/1, EP/L016087/1, the AFOSR COMAS MURI (FA9550-10-1-0558), the ONR NECom MURI, the CIAN NSF ERC under grant EEC-0812072, and TRIF Photonics funding from the state of Arizona and the Micronova, Nanofabrication Centre of Aalto University.


\begin{thebibliography}{99}
%
\bibitem{boyd} Boyd, R. W. {\it Nonlinear Optics}, (Academic Press, 2003).
%
\bibitem{Zipfel_nb_03}
Zipfel, W. R.,  Williams, R. M. \& Webb, W. W.
Nonlinear magic: multiphoton microscopy in the biosciences.
{\it Nat. Biotechnol.}~{\bf 21}, 1369 (2003).
%
\bibitem{Bhawalkar_rpp_96}
Bhawalkar, J. D., He, G. S. \& Prasad,P. N.
Nonlinear multiphoton processes in organic and polymeric materials.
{\it Rep. Prog. Phys.}~{\bf 59}, 1041 (1996).
%

\bibitem{Broderick_JOSAB_02}
Broderick, N. G. R., Bratfalean, R. T., Monro, T. M. Richardson, D. J. \& de Sterke, C. M.
Temperature and wavelength tuning of second-, third-, and fourth-harmonic generation in a two-dimensional hexagonally poled nonlinear crystal.
{\it J. Opt. Soc. Am. B}~{\bf 19}, 2263 (2002).
%
\bibitem{Pavone_13}
Pavone, F. S. \& Campagnola, P.J.
{\it Second Harmonic Generation Imaging.} (CRC Press, 2013).
%
\bibitem{Saleh}
Saleh, B. E. A. \& Teich, M. C.
{\it Fundamentals of Photonics.} (Wiley, 2007).
%
\bibitem{Willner_JLT14}
Willner, A. E., Khaleghi, S., Chitgarha, M. R. \& Yilmaz, O. F.
All-Optical Signal Processing.
{\it J. Lightwave Tech.}~{\bf 32}, 660 (2014).
%

\bibitem{zhu_sci_1997}
Zhu, S. -N., Zhu, Y. -Y. \& Ming, N. -B.,
Quasi-phase-matched third-harmonic generation in a quasi-periodic optical superlattice.
{\it Science}~{\bf 278}, 843 (1997).
%
\bibitem{tsang_pra_1995}
Tsang, T. Y. F.,
Optical third-harmonic generation at interfaces.
{\it Phys. Rev. A}~{\bf 52}, 4116 (1995).
%
\bibitem{Bonaccorso_np_10}
Bonaccorso, F., Sun, Z., Hasan, T. \& Ferrari, A. C.
Graphene photonics and optoelectronics.
{\it Nat. Photon.}~{\bf 4}, 611 (2010).
%
\bibitem{Butler_an_13}
Butler, S. Z. et al.
Progress, challenges, and opportunities in two-dimensional materials beyond graphene.
{\it ACS Nano}~{\bf 7}, 2898-2926 (2013).
%
\bibitem{KoppensNN}
Koppens, F. H. L. et al.
Photodetectors based on graphene, other two-dimensional materials and hybrid systems.
{\it Nat. Nanotechnol.}~{\bf 9}, 780-793 (2014).
%
\bibitem{Ferrari_ns}
Ferrari, A. C. et al.
Science and technology roadmap for graphene, related two-dimensional crystals, and hybrid systems.
{\it Nanoscale}~{\bf 7}, 4598 (2015).
%
\bibitem{Wang_nn_12}
Wang, Q. H.,  Kalantar-Zadeh, K., Kis, A., Coleman, J. N. \&  Strano, M. S.
Electronics and optoelectronics of two-dimensional transition metal dichalcogenides.
{\it Nat. Nanotechnol.}~{\bf 7}, 699-712 (2012).
%
\bibitem{Xu_np_14}
Xu, X., Yao, W., Xiao, D. \& Heinz, T. F.
Spin and pseudospins in layered transition metal dichalcogenides.
{\it Nat. Phys.}~{\bf 10}, 343 (2014).
%
\bibitem{Sun_np_16}
Sun, Z., Martinez, A., \& Wang, F.
Optical modulators with two-dimensional layered materials.
{\it Nat. Photon.}~{\bf 10}, 227-238 (2016).
%
\bibitem{Mak_prl_10}
Mak, F. K., Lee, C.,  Hone, J.,  Shan, J. \& Heinz, T. F.
Atomically thin MoS$_2$: a new direct-gap semiconductor.
{\it Phys. Rev. Lett.}~{\bf 105}, 136805 (2010).
%
\bibitem{Splendiani_nl_2010}
Splendiani, A. et al.
Emerging photoluminescence in monolayer MoS$_2$.
{\it Nano Lett.}~{\bf 10}, 1271 (2010).
%
\bibitem{Eda_nl_2011}
Eda, G., Yamaguchi, H., Voiry, D.,  Fujita, T., Chen, M., \& Chhowalla, M.
Photoluminescence from chemically exfoliated MoS$_2$.
{\it Nano Lett.}~{\bf 11}, 5111 (2011).
%
\bibitem{Goki_an_13}
G. Eda \&  Maier, S. A.
Two-dimensional crystals: managing light for optoelectronics.
{\it ACS Nano}~{\bf 7}, 5660 (2013).
%
\bibitem{berraquero_arxiv_2016}
Berraquero, C. P. et al.
Atomically thin quantum light emitting diodes.
\textit{arXiv:1603.08795} (2016).
%
\bibitem{Amani_sci_15}
 Amani, M. et al.
Near-unity photoluminescence quantum yield in MoS$_2$.
{\it Science}~{\bf 350}, 1065 (2015).
%
\bibitem{Li_nl_13}
Li, Y. et al.
Probing symmetry properties of few-layer MoS$_2$ and h-BN by optical second-harmonic generation.
{\it Nano Letters}~{\bf 13}, 3329 (2013).
%
\bibitem{Kumar_prb_13}
Kumar, N. et al.
Second harmonic microscopy of monolayer MoS$_2$.
{\it Phys. Rev. B}~{\bf 87}, 161403 (2013).
%
\bibitem{Wang_an_13_1}
Wang, K. et al.
Ultrafast saturable absorption of two-dimensional MoS$_2$ nanosheets.
{\it ACS Nano}~{\bf 7}, 9260 (2013).
%
\bibitem{Malard_prb_13}
Malard, L. M.,  Alencar, T.V., Barboza, A. P. M., Mak, K. F. \& de Paula, A. M.
Observation of intense second harmonic generation from MoS$_2$ atomic crystals.
{\it Phys. Rev. B}~{\bf 87}, 201401 (2013).
%
\bibitem{Wang_ami_14}
Wang, R. et al.
Third-harmonic generation in ultrathin films of MoS$_2$.
{\it ACS Appl. Mater. Interfaces}~{\bf 6} 314 (2014).
%
\bibitem{Trolle_prb_14}
Trolle, M. L., Seifert, G. \& Pedersen, T. G.
Theory of excitonic second-harmonic generation in monolayer MoS$_2$.
{\it Phys. Rev. B}~{\bf 89}, 235410 (2014).
%
\bibitem{Clark_prb_14}
Clark, D. et al.
Strong optical nonlinearity of CVD-grown MoS$_2$ monolayer as probed by wavelength-dependent second-harmonic generation.
{\it Phys. Rev. B}~{\bf 90}, 121409 (2014).
%
\bibitem{Bonaccorso_ome_14}
Bonaccorso, F. \& Sun, Z.
Solution processing of graphene, topological insulators and other 2d crystals for ultrafast photonics.
{\it Opt. Mater. Express}~{\bf 4}, 63 (2014).
%
\bibitem{Seyler_nn_15}
Seyler, K. L. et al.
Electrical control of second-harmonic generation in a WSe$_2$ monolayer transistor.
{\it Nat. Nanotechnol.}~{\bf 10}, 407-411 (2015).
%
\bibitem{Kuc_prb_11}
Kuc, A., Zibouche, N. \& Heine, T.
Influence of quantum confinement on the electronic structure of the transition metal sulfide TS$_2$.
{\it Phys. Rev. B}~{\bf 83}, 245213 (2011).
%

%
\bibitem{Shi_prb_13}
Shi, H., Pan, H. , Zhang, Y. -W. \& Yakobson, B. I.
Quasiparticle band structures and optical properties of strained monolayer MoS$_2$ and WS$_2$.
{\it Phys. Rev. B}~{\bf 87}, 155304 (2013).

\bibitem{Kadantsev_ssc_12}
 Kadantsev, E. S. \& Hawrylak, P.
Electronic structure of a single MoS$_2$ monolayer.
{\it Solid State Commun.}~{\bf 152}, 909 (2012).

%
\bibitem{Zahid_aip_13}
Zahid, F.,  Liu, L., Zhu, Y.,  Wang, J. \&  Guo, H.
A generic tight-binding model for monolayer, bilayer and bulk MoS$_2$.
{\it AIP Advances}~{\bf 3}, 052111 (2013).
%
\bibitem{Kormanyos_prb_13}
Korm\'anyos, A. et al.
Monolayer MoS$_{2}$: Trigonal warping, the $\ensuremath{\Gamma}$ valley, and spin-orbit coupling effects.
{\it Phys. Rev. B}~{\bf 88}, 045416 (2013).
%
\bibitem{Qiu_prl_13}
Y. Qiu, D.,  da Jornada. F.H. \&   Louie, S. G.
Optical Spectrum of MoS$_2$: Many-Body Effects and Diversity of Exciton States
{\it Phys. Rev. lett.}~{\bf111}, 216805 (2013).
%
\bibitem{gibertini_prb_2014}
Gibertini, M., Pellegrino, F. M. D., Marzari, N. \& Polini, M.
Spin-resolved optical conductivity of two-dimensional group-VIB transition-metal dichalcogenides.
{\it Phys. Rev. B}~{\bf 90}, 245411 (2014).
%
\bibitem{rostami_prb_2016}
Rostami, H. \& Polini, M.
Theory of third harmonic generation in graphene: a diagrammatic approach.
{\it Phys. Rev. B}~{\bf 93}, 161411 (2016).
%
\bibitem{RG15}
Rostami, H., Rold\'an, R., Cappelluti, E., Asgari, R. \& Guinea, F.
Theory of strain in single-layer transition metal dichalcogenides.
{\it Phys. Rev. B}~{\bf 92}, 195402 (2015).
%
\bibitem{Novoselov_PNAS05}
Novoselov, K. S. et al.
Two-dimensional atomic crystals.
{\it Proc. Natl. Acad. Sci. USA} {\bf 102}, 10451 (2005).
%
\bibitem{Bonaccorso_mt_12}
Bonaccorso, F. et al.
Production and processing of graphene and 2d crystals.
{\it Mater. Today}~{\bf 15}, 564 (2012).
%
\bibitem{SundNL13}
Sundaram, R. S. et al.
Electroluminescence in single layer MoS$_2$.
{\it Nano Lett.}~{\bf 13}, 1416 (2013).
%
\bibitem{Casiraghi_NL07}
Casiraghi, C. et al.
Rayleigh imaging of graphene and graphene layers.
 {\it Nano Lett.}~{\bf 7}, 2711 (2007).
%
\bibitem{LeeACSN4}
Lee, J., Novoselov, K. S. \& Shin, H. S.
Interaction between metal and graphene: dependence on the layer number of graphene.
{\it ACS Nano}~{\bf 4}, 2695 (2010).
%
\bibitem{ZhangPRB87}
Zhang, X. et al.
Raman spectroscopy of shear and layer breathing modes in multilayer MoS$_2$.
{\it Phys. Rev. B}~{\bf 87}, 115413 (2013).
%
\bibitem{saynatjoki13}
S\"ayn\"atjoki, A. et al.
Rapid large-area multiphoton microscopy for characterization of graphene.
{\it ACS Nano}~{\bf 7}, 8441 (2013).
%
\bibitem{kieu10}
Kieu, K, Jones, R. \& Peyghambarian, N.
Generation of few-cycle pulses from an amplified carbon nanotube mode-locked fiber laser system.
{\it IEEE Photon. Technol. Lett.}~{\bf 22}, 1521 (2010).
%

\bibitem{kieu10high}
Kieu, K, Jones, R. \& Peyghambarian, N.
High power femtosecond source near 1 micron based on an all-fiber Er-doped mode-locked laser.
{\it Opt. Express}~{\bf 18}, 21350 (2010).
%

\bibitem{Janisch_Scirep_14}
Janisch, C. et al.
Extraordinary second harmonic generation in tungsten disulfide monolayers.
{\it Sci. Rep.}~{\bf 4}, 5530 (2014).
%
\bibitem{hendry_prl_2010}
Hendry, E., Hale, P. J., Moger, J.,  Savchenko, A. K. \& Mikhailov, S. A.
Coherent nonlinear optical response of graphene.
{\it Phys. Rev. Lett.}~{\bf 105}, 097401 (2010).
%
\bibitem{kumar_prb_2013}
Kumar, N. et al.
Third harmonic generation in graphene and few-layer graphite films.
{\it Phys. Rev. B}~{\bf 87}, 121406(R) (2013).
%
\bibitem{hong_prx_2013}
Hong, S. -Y. et al.
Optical third-harmonic generation in graphene.
{\it Phys. Rev. X}~{\bf 3}, 021014 (2013).
%

\bibitem{Nair_s_08}
Nair, R. R.; Blake, P.; Grigorenko, A. N.; Novoselov, K. S.; Booth, T. J.; Stauber, T.; Peres, N. M. R.; Geim, A. K. Fine Structure Constant Defines Visual Transparency of Graphene. {\it Science} {\bf 320}, 1308-1308 (2008).

\bibitem{Gruning_prb_14}
Gr{\"u}ning, M. \& Attaccalite, C.
Second harmonic generation in h-BN and MoS$_2$ monolayers: Role of electron-hole interaction.
{\it Phys. Rev. B}~{\bf 89}, 081102 (2014).

\bibitem{pfister_ol_1997}
Pfister, O. et al.
Continuous-wave frequency tripling and quadrupling by simultaneous three-wave mixings in periodically poled crystals:application to a two-step 1.19-10.71-$\mu$m frequency bridge.
{\it Opt. Lett.}~{\bf 22}, 1211 (1997).
 %
 \bibitem{Butcher_and_Cotter}
Butcher, P. N.  \&  Cotter, D.
{\it The elements of nonlinear optics} (Cambridge University Press, 1990).
%
\bibitem{Radisavljevic_nn_11}
Radisavljevic, B., Radenovic, A., Brivio, J., Giacometti, V. \& Kis, A.
Single-layer MoS$_2$ transistors.
{\it Nat. Nanotechnol.} ~{\bf 6}, 147-150 (2011).
%

\bibitem{RG15_2}
Rostami, H., Asgari, R. \& Guinea, F.
Edge modes in zigzag and armchair ribbons of monolayer MoS$_2$.
{\it arXiv:1511.07003} (2015).

\bibitem{RMA13}
Rostami, H., Moghaddam, A. G. \& Asgari, R.
Effective lattice Hamiltonian for monolayer MoS$_2$: Tailoring electronic structure with perpendicular electric and magnetic fields.
{\it Phys. Rev. B}~{\bf 88}, 085440 (2013).

\bibitem{AH14}
Alidoust, N. et al.
Observation of monolayer valence band spin-orbit effect and induced quantum well states in ${\rm MoX}_2$.
{\it Nature Commun}.~{\bf 5} 4673 (2014).

\bibitem{Giuliani_and_Vignale}
Giuliani, G. F. \& Vignale, G.
{\it Quantum Theory of the Electron Liquid} (Cambridge University Press, 2005).

\bibitem{RKP16}
Rostami, H., Katsnelson, M.K. \& Polini, M.
Gauge invariance in second-order nonlinear optics.
(in preparation).

\bibitem{Griffiths}
Griffiths, D. J., {\it Introduction to Electrodynamics} (Pearson,  2012).

\bibitem{Mahan}
Mahan, G. D.
{\it Many-Particle Physics: Physics of Solids and Liquids} (Springer US, 1981).


\bibitem{J12}
Jafari, S. A.
Nonlinear optical response in gapped graphene.
{\it J. Phys.: Condens. Matter}~{\bf 24}, 205802 (2012).


\bibitem{chirolli_prl_2012}
Chirolli, L., Polini, M., Giovannetti, V. \& MacDonald, A. H.
Drude weight, cyclotron resonance, and the Dicke model of graphene cavity QED.
{\it Phys. Rev. Lett.}~{\bf 109}, 267404 (2012).

\bibitem{MMG13}
Margulis, Vl. A., Muryumin, E. E. \& Gaiduk, E. A.
Optical second-harmonic generation from two-dimensional hexagonal crystals with broken space inversion symmetry.
{\it J. Phys.: Condens. Matter}~{\bf 25}, 195302 (2013).


\bibitem{WX12}
Wu, S., Mao, L., Jones, A. M., Yao, W., Zhang, C. \& Xu, X.
Quantum-enhanced tunable second-order optical nonlinearity in bilayer graphene.
{\it Nano Lett.}~{\bf 12}, 2032 (2012).


\bibitem{BP15}
Brun, S. J., \& Pedersen, T. G.
Intense and tunable second-harmonic generation in biased bilayer graphene.
{\it Phys. Rev. B}~{\bf 91}, 205405 (2015).

\bibitem{CG13}
Cappelluti, E., Rold\'an, R., Silva-Guill\'en, J. A., Ordej\'on, P. \& Guinea, F.
Tight-binding model and direct-gap/indirect-gap transition in single-layer and multilayer MoS$_2$.
{\it Phys. Rev. B}~{\bf 88}, 075409 (2013).

\bibitem{RO14}
Rold\'an, R. et al.
Momentum dependence of spin-orbit interaction effects in single-layer and multi-layer transition metal dichalcogenides.
{\it 2D Mater.}~{\bf 1}, 034003 (2014).

\end{thebibliography}
\end{document}